%% file: main.tex
\documentclass{article}

\input{header/preamble}

\input{header/math}
\input{header/defns}

\usepackage[accepted]{icml2020}

\begin{document}
\icmltitlerunning{Jukebox: A Generative Model for Music}

\twocolumn[
\icmltitle{Jukebox: A Generative Model for Music}



\icmlsetsymbol{equal}{*}

\begin{icmlauthorlist}
\icmlauthor{Prafulla Dhariwal}{equal,openai}
\icmlauthor{Heewoo Jun}{equal,openai}
\icmlauthor{Christine Payne}{equal,openai}
\icmlauthor{Jong Wook Kim}{openai}
\icmlauthor{Alec Radford}{openai}
\icmlauthor{Ilya Sutskever}{openai}
\end{icmlauthorlist}

\icmlaffiliation{openai}{OpenAI, San Francisco}

\icmlcorrespondingauthor{}{jukebox@openai.com}

\icmlkeywords{Machine Learning, Generative Models, Music}

\vskip 0.3in
]



\printAffiliationsAndNotice{\icmlEqualContribution} 

\begin{abstract}
We introduce Jukebox, a model that generates music with singing in the raw audio domain. We tackle the long context of raw audio using a multi-scale VQ-VAE to compress it to discrete codes, and modeling those using autoregressive Transformers. We show that the combined model at scale can generate high-fidelity and diverse songs with coherence up to multiple minutes. We can condition on artist and genre to steer the musical and vocal style, and on unaligned lyrics to make the singing more controllable. We are releasing thousands of non cherry-picked \href{https://jukebox.openai.com}{samples}, along with model weights and  \href{https://github.com/openai/jukebox}{code}.
\end{abstract}

\section{Introduction}
Music is an integral part of human culture, existing from the earliest periods of human civilization and evolving into a wide diversity of forms. 
It evokes a unique human spirit in its creation, and the question of whether computers can ever capture this creative process has fascinated computer scientists for decades. 
We have had algorithms generating piano sheet music \cite{hiller,rulebased,deepbach,coconet}, digital vocoders generating a singer’s voice \cite{concat-singing-synth,hmm-singing-1,neural-singing-synth} and also synthesizers producing timbres for various musical instruments
\cite{nsynth,gansynth}.
Each captures a specific aspect of music generation: melody, composition, timbre, and the human voice singing. However, a single system to do it all remains elusive.

The field of generative models has made tremendous progress in the last few years. One of the aims of generative modeling is to capture the salient aspects of the data and to generate new instances indistinguishable from the true data 
The hypothesis is that by learning to produce the data we can learn the best features of the data$^1$. We are surrounded by highly complex distributions in the visual, audio, and text domain, and in recent years we have developed advances in text generation \cite{gpt2}, speech generation \cite{reswavenet} and image generation \cite{biggan,vqvae2}. The rate of progress in this field has been rapid, where only a few years ago we had algorithms producing blurry faces \cite{vae,gan} but now we now can generate high-resolution faces indistinguishable from real ones \cite{stylegan}. \hiddenfootnote{$^1$Richard Feynmann famously said, ``What I cannot create, I do not understand''}

Generative models have been applied to the music generation task too. Earlier models generated music symbolically in the form of a pianoroll, which specifies the timing, pitch, velocity, and instrument of each note to be played. 
\cite{midinet,musegan,musictransformer,musenet,musicvae,hrnn}. The symbolic approach makes the modeling problem easier by working on the problem in the lower-dimensional space. However, it constrains the music that can be generated to being a specific sequence of notes and a fixed set of instruments to render with. In parallel, researchers have been pursuing the non-symbolic approach, where they try to produce music directly as a piece of audio. This makes the problem more challenging, as the space of raw audio is extremely high dimensional with a high amount of information content to model. There has been some success, with models producing piano pieces either in the raw audio domain \cite{wavenet,samplernn,parallelwavegan} or in the spectrogram domain \cite{melnet}. 
The key bottleneck is that modeling the raw audio directly introduces extremely long-range dependencies, making it computationally challenging to learn the high-level semantics of music. 
A way to reduce 
the difficulty is to learn a lower-dimensional encoding of the audio with the goal of losing the less important information but retaining most of the musical information. This approach has demonstrated some success in generating short instrumental pieces restricted to a set of a few instruments \cite{vqvae, sanderschallenge}.

In this work, we show that we can use state-of-the-art deep generative models to produce a single system capable of generating diverse high-fidelity music in the raw audio domain, with long-range coherence spanning multiple minutes. Our approach uses a hierarchical VQ-VAE architecture \cite{vqvae2} to compress audio into a discrete space, with a loss function designed to retain the maximum amount of musical information, while doing so at increasing levels of compression. We use an autoregressive Sparse Transformer \cite{sparsetransformer, aiayn} trained with maximum-likelihood estimation over this compressed space, and also train autoregressive upsamplers to recreate the lost information at each level of compression. 

We show that our models can produce songs from highly diverse genres of music like rock, hip-hop, and jazz. They can capture melody, rhythm, long-range composition, and timbres for a wide variety of instruments, as well as the styles and voices of singers to be produced with the music. We can also generate novel completions of existing songs. Our approach allows the option to influence the generation process: by swapping the top prior with a conditional prior, we can condition on lyrics to tell the singer what to sing, or on midi to control the composition. We release our model weights and training and sampling code at \href{https://github.com/openai/jukebox}{https://github.com/openai/jukebox}.

\section{Background}

We consider music in the raw audio domain represented as a continuous waveform $\x \in [-1,1]^T$, where the number of samples $T$ is the product of the audio duration and the sampling rate typically ranging from 16 kHz to 48 kHz.
For music, CD quality audio, 44.1 kHz samples stored in 16 bit precision, is typically enough to capture the range of frequencies perceptible to humans. As an example, a four-minute-long audio segment will have an input length of ${\sim}10$ million, where each position can have 16 bits of information. In comparison, a high-resolution RGB image with $1024\times1024$ pixels has an input length of ${\sim}3$ million, and each position has 24 bits of information. This makes learning a generative model for music extremely computationally demanding with increasingly longer durations; we have to capture a wide range of musical structures from timbre to global coherence while simultaneously modeling a large amount of diversity.

\subsection{VQ-VAE}  

To make this task feasible, we use the VQ-VAE \cite{vqvae,sanderschallenge,vqvae2} to compress raw audio to a lower-dimensional space. A one-dimensional VQ-VAE learns to encode an input sequence
$\x = \langle \x_t \rangle_{t = 1}^T$
using
a sequence of discrete tokens
$\z = \langle z_s \in [K] \rangle_{s = 1}^S$, 
where $K$ denotes the vocabulary size and we call the ratio $T / S$ the hop length. It consists of an encoder $E(\x)$ which encodes $\x$ into a sequence of 
latent vectors $\h = \langle \h_s \rangle_{s = 1}^S$,
a bottleneck that quantizes $\h_s \mapsto \e_{z_s}$ by mapping each $\h_s$ to its nearest vector $\e_{z_s}$ from a codebook 
$C = \{ \e_k \}_{k=1}^K$,
and a decoder $D(\e)$ that decodes the embedding vectors back to the input space. It is thus an auto-encoder with a discretization bottleneck. The VQ-VAE is trained using the following objective:
\begin{align}
\Loss ~&=~ \Loss_\recons + \Loss_\codebook +  \beta\Loss_\commit \label{eq:loss:vq} \\[0.4em]
\Loss_\recons ~&=~ \tfrac{1}{T}\smallsum_t\, \lVert \x_t - D(\e_{z_t}) \rVert_2^2 \label{eq:loss:recons}\\
\Loss_\codebook ~&=~ \tfrac{1}{S}\smallsum_s\, \lVert \sg \left [ \h_s \right ] - \e_{z_s}
\label{eq:loss:codebook}
\rVert_2^2 \\
\Loss_\commit ~&=~ \tfrac{1}{S}\smallsum_s\, \lVert \h_s - \sg \left [ \e_{z_s} \right ] \rVert_2^2
\label{eq:loss:commit}
\end{align}
where $\sg$ denotes the stop-gradient operation, which passes zero gradient during backpropagation.
The reconstruction loss $\Loss_\recons$ penalizes for the distance between the input $\x$ and the reconstructed output $\widehat{\x} = D(\e_{z})$, and $\Loss_\codebook$ penalizes the codebook for the distance between the encodings $\h$ and their nearest neighbors $\e_{z}$ from the codebook. 
To stabilize the encoder, we also add $\Loss_\commit$ to prevent the encodings from fluctuating too much, where the weight $\beta$ controls the amount of contribution of this loss.
To speed~up training, the codebook loss $\Loss_\codebook$ instead uses EMA updates over the codebook variables. 
\citet{vqvae2} extends this to a hierarchical model where they train a single encoder and decoder but break up the latent sequence $\h$ into a multi-level representation $[\h^{(1)}, \cdots, \h^{(L)}]$ with decreasing sequence lengths, each learning its own codebook $C^{(l)}$.
They use non-autoregressive encoder-decoders and jointly train all levels with a simple mean-squared loss.

\begin{figure*}[ht]
    \centering
    \includegraphics[width=\textwidth]{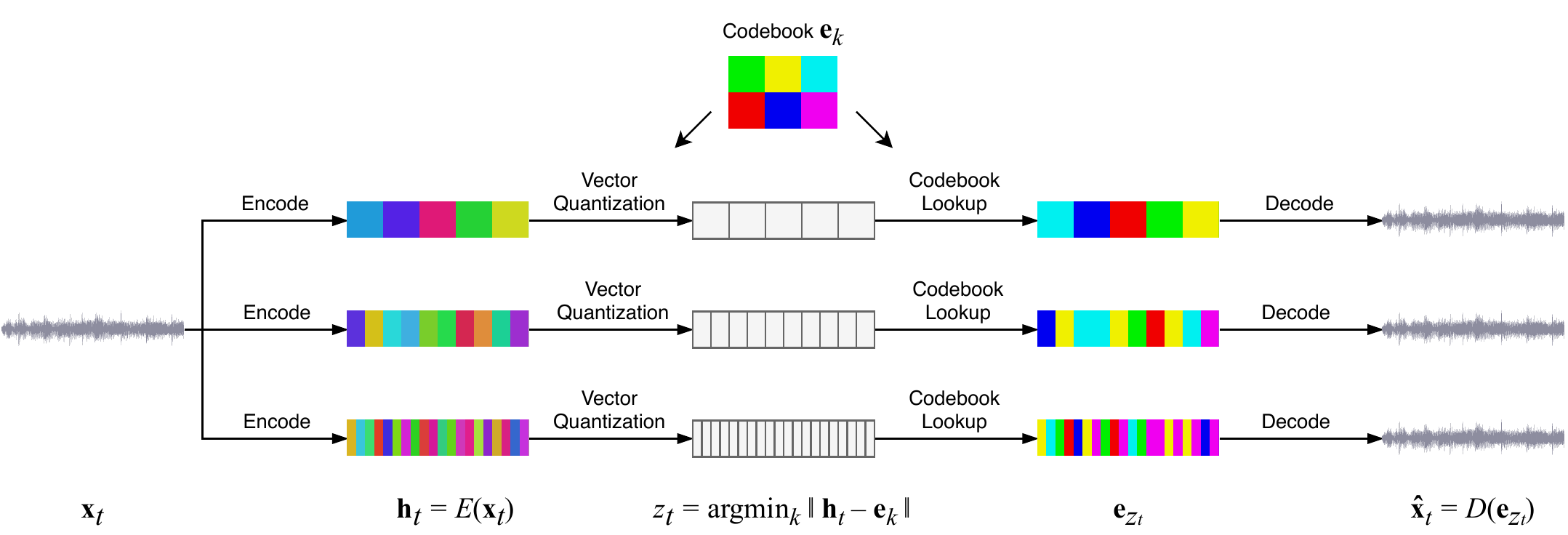}
    \caption{We first train three separate VQ-VAE models with different temporal resolutions. At each level, the input audio is segmented and encoded into latent vectors $\mathbf{h}_t$, which are then quantized to the closest codebook vectors $\mathbf{e}_{z_t}$. The code $z_t$ is a discrete representation of the audio that we later train our prior on. The decoder takes the sequence of codebook vectors and reconstructs the audio. The top level learns the highest degree of abstraction, since it is encoding longer audio per token while keeping the codebook size the same. Audio can be reconstructed using the codes at any one of the abstraction levels, where the least abstract bottom-level codes result in the highest-quality audio, as shown in Figure \ref{fig:vq:spectrogram}.
    For the detailed structure of each component, see Figure \ref{fig:architecture:vqvae-components}.
    }
    \label{fig:architecture:vqvae}
\end{figure*}

\section{Music VQ-VAE}

Inspired by the results from the hierarchical VQ-VAE model \cite{vqvae2} for images, we consider applying the same technique to model raw audio using three different levels of abstraction, as illustrated in Figure \ref{fig:architecture:vqvae}.
At each level, we use residual networks consisting of WaveNet-style non-causal 1-D dilated convolutions, interleaved with downsampling and upsampling 1-D convolutions to match different hop lengths.
A detailed description of the architecture is provided in Appendix \ref{sec:music-vqvae}.
We make a number of modifications to our VQ-VAE compared to the ones in \cite{vqvae,vqvae2}, as described in the following subsections.

\subsection{Random restarts for embeddings}
\label{randomrestart}

VQ-VAEs are known to suffer from codebook collapse, wherein all encodings get mapped to a single or few embedding vectors while the other embedding vectors in the codebook are not used, reducing the information capacity of the bottleneck. To prevent this, we use random restarts: when the mean usage of a codebook vector falls below a threshold, we randomly reset it to one of the encoder outputs from the current batch. This ensures all vectors in the codebook are being used and thus have a gradient to learn from, mitigating codebook collapse.

\subsection{Separated Autoencoders}
\label{sec:separate-autoencoder}

When using the hierarchical VQ-VAE from \cite{vqvae2} for raw audio, we observed that the bottlenecked top level is utilized very little and sometimes experiences a complete collapse, as the model decides to pass all information through the less bottlenecked lower levels.
To maximize the amount of information stored at each level, we simply train separate autoencoders with varying hop lengths. Discrete codes from each level can be treated as independent encodings of the input at different levels of compression.

\subsection{Spectral Loss}\label{sec:spectral-loss}
When using only the sample-level reconstruction loss, the model learns to reconstruct low frequencies only. To capture mid-to-high frequencies,
we add a spectral loss which is defined as 
$$\Loss_\spec = \Norm{ \abs{\stft(\x)} - \abs{\stft(\widehat \x)} }_2$$
It encourages the model to match the spectral components without paying attention to phase which 
is more difficult to learn.
This is similar to the use of power loss \cite{parallelwavenet} and spectral convergence \cite{mcnn} when training parallel decoders for raw audio. 
One difference between the latter approach and ours is that we are no longer optimizing the spectral signal-to-noise ratio; dividing by the magnitude of the signal results in numerical instability for mostly silent inputs.
To prevent the model from overfitting to a particular choice of the STFT parameters, we use the sum of the spectral losses $\Loss_\spec$ calculated over multiple STFT parameters that trade-off time and frequency resolutions \cite{parallelwavegan}.

\section{Music Priors and Upsamplers}

After training the VQ-VAE, we need to learn a prior $p(\z)$ over the compressed space to generate samples. We break up the prior model as 
\begin{align}
p(\z) &= p(\z^\text{top}, \z^\text{middle}, \z^\text{bottom})  \\
&= p(\z^\text{top}) p(\z^\text{middle}|\z^\text{top}) p(\z^\text{bottom}|\z^\text{middle},\z^\text{top})\label{eqn:prior}
\end{align}
and train separate models for the top-level prior $p(\z^\text{top})$, and upsamplers $p(\z^\text{middle}|\z^\text{top})$ and $p(\z^\text{bottom}|\z^\text{middle},\z^\text{top})$. Each of these is an autoregressive modeling problem in the discrete token space produced by the VQ-VAE. We use Transformers with sparse attention \cite{aiayn,sparsetransformer} as they are currently the SOTA in autoregressive modeling. We propose a simplified version which we call the Scalable Transformer, that is easier to implement and scale (see Appendix \ref{scalabletransformer} for details). 

For the upsamplers, we need to provide the autoregressive Transformers with conditioning information from the codes of the upper levels. To do so, we use a deep residual WaveNet \cite{reswavenet} followed by an upsampling strided convolution and a layer norm \cite{layernorm}, and add the output as extra positional information to the embeddings of the current level. We condition the lower levels only on the chunk of upper level codes that correspond to the same segment of raw audio. 

At each level, we use Transformers over the same context length of discrete codes, which correspond to increasing the raw audio length with larger hop lengths, and modeling longer temporal dependencies at the higher levels while keeping the same computational footprint for training each level. As our VQ-VAE is convolutional, we can use the same VQ-VAE to produce codes for arbitrary lengths of audio.

\begin{figure}[ht!]
    \begin{subfigure}{\columnwidth}
    \centering
    \includegraphics[width=\columnwidth]{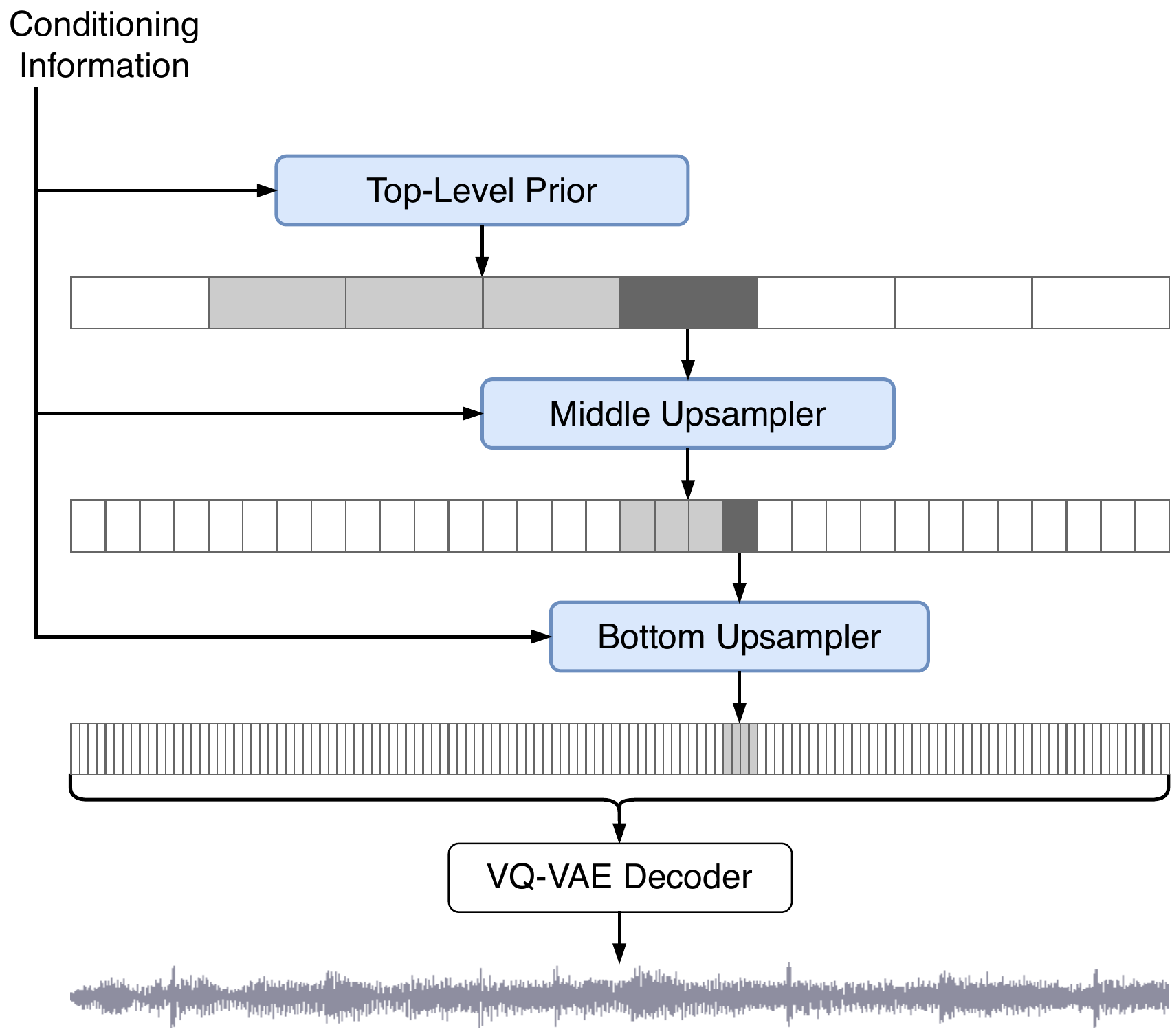}
    \caption{\textbf{Ancestral sampling}: Priors for the VQ-VAE codes are trained using a cascade of Transformer models, shown in blue. Each model takes conditioning information such as genre, artist, timing, and lyrics, and the upsampler models are also conditioned on the codes from the upper levels. To generate music, the VQ-VAE codes are sampled from top to bottom using the conditioning information for control, after which the VQ-VAE decoder can convert the bottom-level codes to audio.}
    \label{fig:architecture:ancestral}
    \end{subfigure}\\[2.5em]
    \begin{subfigure}{\columnwidth}
    \centering
    \includegraphics[width=\columnwidth]{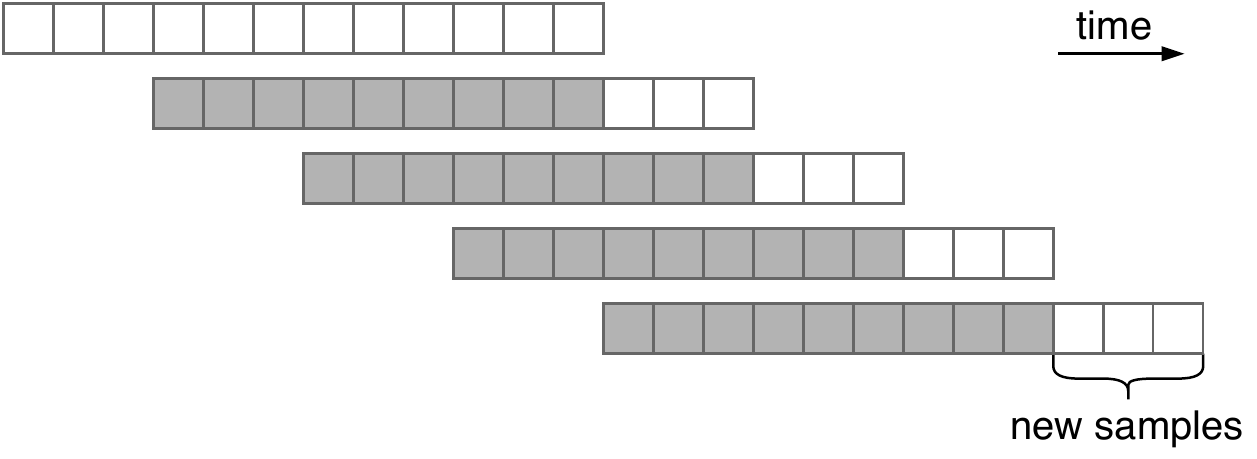}
    \caption{\textbf{Windowed sampling}: To generate music longer than the model's context length (12 in this figure), we repeatedly sample continuations at each level, using overlapping windows of previous codes as the context. The overlap amount is a hyperparameter, and the figure shows an example of 75\% overlap with hop length 3.}
    \label{fig:architecture:windowed}
    \end{subfigure}\\[2.5em]
    \begin{subfigure}{\columnwidth}
    \centering
    \includegraphics[width=\columnwidth]{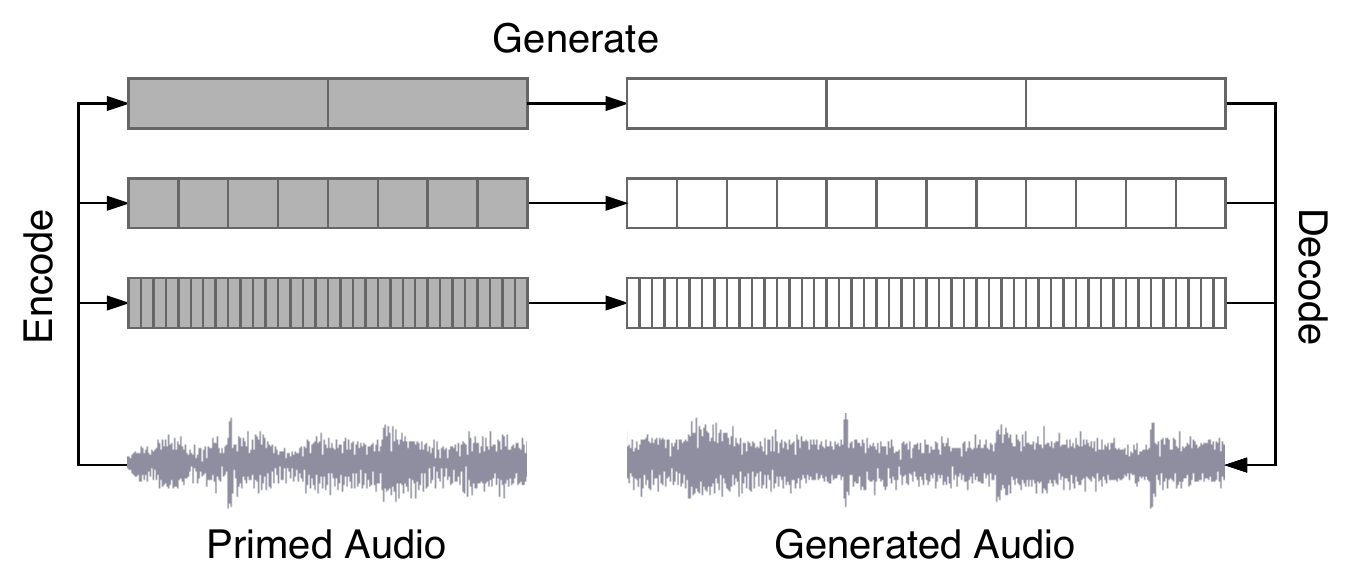}
    \caption{\textbf{Primed sampling}: The model can generate continuations of an existing audio signal by converting it into the VQ-VAE codes and sampling the subsequent codes in each level.}
    \label{fig:architecture:primed}
    \end{subfigure}
    \caption{Sampling methods for generating music}   \label{fig:architecture:generation}
    \vspace{-2em}
\end{figure}


\subsection{Artist, Genre, and Timing Conditioning}
Our generative model can be made more controllable by providing additional conditioning signals while training. For our first models, we provide artist and genre labels for the songs. This has two advantages: first, it reduces the entropy of the audio prediction, so the model is able to achieve better quality in any particular style. Second, at generation time, we are able to steer the model to generate in a style of our choosing. Additionally, we attach a timing signal for each segment at training time. This signal includes the total duration of the piece, the start time of that particular sample and how much fraction of the song that has elapsed.  This allows the model to learn audio patterns that depend on the overall structure, such as spoken or instrumental introductions and applause at the end of a piece. 

\subsection{Lyrics Conditioning}
While the conditioned models above are able to generate songs of diverse genres and artistic styles, singing voices generated by those models, while often sung in a compelling melody, are mostly composed of babbling, rarely producing recognizable English words. In order to be able to control the generative model with lyrics, we provide more context at training time by conditioning the model on the lyrics corresponding to each audio segment, allowing the model to produce singing simultaneosly with the music. 

{\bfseries Lyrics-to-singing (LTS) task}: The conditioning signal only includes the text of the lyrics, without timing or vocalisation information. We thus have to model the temporal alignment of lyrics and singing, the artists voice and also the diversity of ways one can sing a phrase depending on the pitch, melody, rhythm and even genre of the song. The conditioning data isn't precise as the lyrics data often contains textual references to repeated sections like ``chorus'' or mismatching portions of lyrics with the corresponding music. There is also no separation between lead vocals, accompanying vocals and the background music in target audio. This makes the Lyrics-to-singing (LTS) task significantly more challenging than the corresponding Text-to-speech (TTS) task.

{\bfseries Providing lyrics for chunks of audio}: Our dataset includes song-level lyrics, but to make the task easier we train on shorter (24 sec) chunks of audio. To provide the lyrics corresponding to the audio during training, we began with a simple heuristics of aligning the characters of the lyrics to linearly span the duration of each song, and pass a fixed-side window of characters centered around the current segment during training. While this simple strategy of linear alignment worked surprisingly well, we found that it fails for certain genres such as hip-hop with fast lyrics. To address this, we use Spleeter \cite{spleeter} to extract vocals from each song and run NUS AutoLyricsAlign \cite{autolyricsalign} on the extracted vocals to obtain a word-level alignments of the lyrics, allowing us to more accurately provide the lyrics for a given chunk of audio. We choose a large enough window so that the actual lyrics have a high probability of being inside the window. 

\begin{figure}[t]
\centering
\includegraphics[width=\columnwidth]{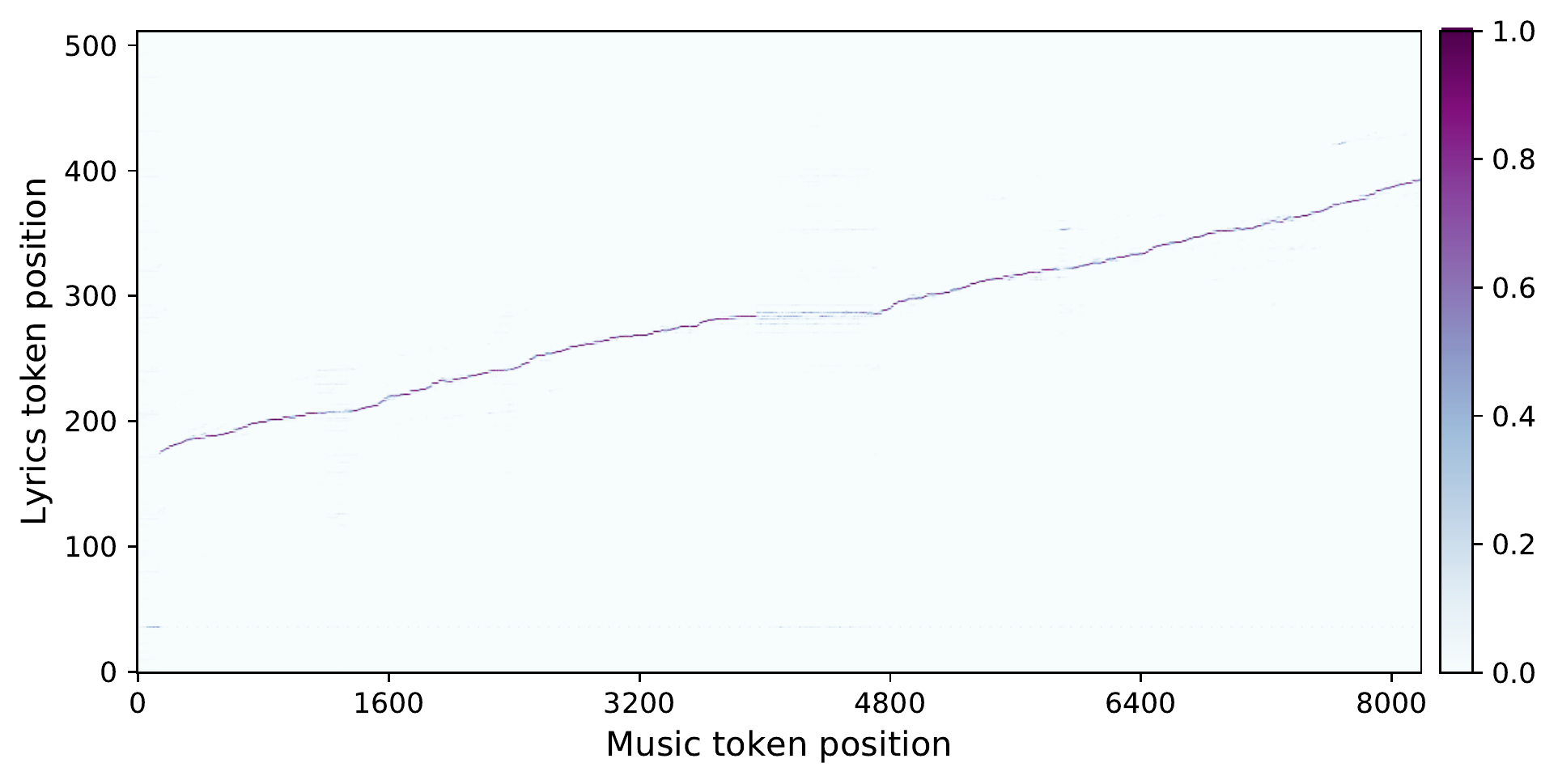}
\caption{Lyrics-singing alignment learned by one of the encoder-decoder attention layers. The $x$-axis is the position of music queries, and the $y$-axis is the position of lyric keys. The positions attended to by the decoder correspond to the characters being sung.}
\label{fig:alignment}
\end{figure}

{\bfseries Encoder-decoder model}: We use an encoder-decoder style model to condition on the characters of the lyrics, with the encoder producing features from the lyrics which are attended to by the decoder which produces the top level music tokens. The lyrics encoder is a Transformer with an autoregressive modeling loss for lyrics, and its last level is used as features of the lyrics. 
In the music decoder, we interleave a few additional layers with encoder-decoder attention where the queries from the music tokens are only allowed to attend to keys and values from the lyrics tokens. These layers attend on the activation from the last layer of the lyrics encoder (see Figure \ref{fig:architecture:encdec}). In Figure \ref{fig:alignment}, we see that the attention pattern learned by one of these layers corresponds to the alignment between the lyrics and the singing.

\subsection{Decoder Pretraining}
\label{sec:surgery}
To reduce computation required to train the lyrics conditional model, we use a pretrained unconditional top-level prior as our decoder and introduce the lyrics encoder using model surgery \cite{dota2}. We initialize the output projection weights in the MLP and the attention layers of these residual blocks to zeros \cite{fixup}, so that the added layers perform the identity function at initialization. Thus, at initialization the model behaves identically as the pretrained decoder, but there is still a gradient with respect to the encoder state and parameters\footnote{The gradient also needs to break symmetry with the encoder output features, which is the case here since the weights of the input projections in the attention are not zero.}, allowing the model to learn to use the encoder. 

\subsection{Sampling}
After we have trained our VQ-VAE, upsamplers, and top level priors, we can then use them to sample novel songs. 

{\bfseries Ancestral sampling}: We first generate the top level codes one token at a time by the usual ancestral sampling process (see Figure \ref{fig:architecture:ancestral}): generating the first token, then passing all previously generated tokens into the model as inputs and outputting the next token conditioned on all previous tokens. We then run our conditioning wavenet on the top level codes to produce the conditioning information for the middle level and sample ancestrally from it too, and do the same for the bottom level. 

{\bfseries Windowed sampling}: To sample segments longer than the context length, we use windowed sampling, where we move ahead our sampling window by half our context and continue sampling conditioned on this half context (See Figure \ref{fig:architecture:windowed}). We can trade off speed for quality by using a smaller hop length here.

{\bfseries Primed sampling}: Instead of sampling the entire token sequence from the model, we can also run a forward pass of the VQ-VAE to obtain the top, middle, and bottom level codes corresponding to a segment from an actual song, as shown in Figure \ref{fig:architecture:primed}. We can use these as the initial tokens in our ancestral sampling process and continue sampling from these to produce novel completions of the song.

\section{Experiments}


\subsection{Dataset}
We scraped a new dataset of 1.2 million songs (600k of which in English), paired with the lyrics and metadata from LyricWiki \cite{lyricwiki}. The metadata includes artist, album, genre, and year of the release, along with common moods or playlist keywords associated with each song. 
We train on 32 bit, 44.1 kHz raw audio and perform data augmentation by randomly downmixing the right and left channels to produce mono channel audio.

\subsection{Training Details}
For the music VQ-VAE, we use 3 levels of bottlenecks compressing 44 kHz audio in dimensionality by 8x, 32x, and 128x respectively, with a codebook size of 2048 for each level. 
The VQ-VAE has 2 million parameters and is trained on 9-second audio clips on 256 V100 for 3 days. We used exponential moving average to update the codebook following \citet{vqvae2}.
For our prior and upsampler models, we use a context of 8192 tokens of VQ-VAE codes, which corresponds to approximately 24, 6, and 1.5 seconds of raw audio at the top, middle, and bottom level, respectively. The upsamplers have one billion parameters and are trained on 128 V100s for 2 weeks, and the top-level prior has 5 billion parameters and is trained on 512 V100s for 4 weeks. We use Adam with learning rate $0.00015$ and weight decay of $0.002$. For lyrics conditioning, we reuse the prior and add a small encoder, after which we train the model on 512 V100s for 2 weeks. 
The detailed hyperparameters for our models and training are provided in Appendix \ref{sec:hps}. 

\subsection{Samples}

We trained a sequence of models with increasing sample quality. Our first model was trained on the MAESTRO dataset using 22 kHz VQ-VAE codes and relatively small prior models. We observed that this could generate high fidelity \href{https://soundcloud.com/openai_audio/sets/timeline-august/s-VOsPA}{classical music} samples with piano and occasional violin. We then collected a larger and more diverse dataset of songs with genre and artist labels. The same model when trained on this new dataset was able to produce \href{https://soundcloud.com/openai_audio/sets/timeline-october/s-bLoQj}{diverse samples} other than classical music, and demonstrated musicality and coherence over more than a minute.

Despite the novelty of being able to generate generally high fidelity and coherent songs, sample quality was still limited by a number of factors. First, the use of 22 kHz sampling rate along with small upsamplers introduced noise both in the upsampling and decoding steps, which we hear as grainy texture. We improved fidelity by using 44 kHz VQ-VAE and 1B parameter upsamplers in all subsequent experiments at the expense of longer rendering time.

Second, the 1B top-level prior was not big enough to produce singing and diverse musical timbres. We first explored increasing the model size to 5 billion parameters. Larger capacity allowed better modeling of the broader distribution of songs, resulting in samples with \href{https://soundcloud.com/openai_audio/sets/timeline-january}{better musicality, longer coherence and initial singing}. While there is an overall qualitative improvement, the unconditional model still struggled to sing recognizable words. Training a seq2seq model with lyric conditioning and limiting the dataset only to songs primarily in English made \href{https://soundcloud.com/openai_audio/sets/jukebox/s-P9Egj}{singing} both intelligible and controllable.

The final model, which we call Jukebox, uses all these improvements. Because everyone experiences music differently, it is generally tricky and not very meaningful to evaluate samples by the mean opinion score or FID-like metrics. We manually evaluate coherence, musicality, diversity, and novelty of generated samples.
The links to curated examples are embedded in text.

{\bfseries \href{https://soundcloud.com/openai_audio/sets/jukebox-samples-coherence}{Coherence}:} We find the samples stay very coherent musically through the context length of the top-level prior (approximately 24 seconds), and they maintain similar harmonies and textures as we slide the window to generate longer samples.  However, because the top-level does not have the context of the entire song, we do not hear long term musical patterns, and we would never hear choruses or melodies that repeat.

The generations progress through beginnings of songs (for example applause or slow instrumental warm-ups), through sections that sound chorus-like, through instrumental interludes, and then fading or otherwise wrapping up at the end. The top-level prior always knows what fraction of the song is complete time-wise, so it is able to imitate appropriate beginnings, middles and ends.   

{\bfseries \href {https://soundcloud.com/openai_audio/sets/jukebox-samples-musicality}{Musicality}:} The samples frequently imitate familiar musical harmonies and the lyrics are usually set in ways that are very natural. Frequently the highest or longest notes of the melody match words that a human singer would choose to emphasize, and the lyrics are almost always rendered in ways that capture the prosody of the phrases.  This is noticeable in hip hop generations, where the model reliably captures the rhythm of spoken text. We do find that the generated melodies are usually less interesting than human composed melodies. In particular, we do not hear the antecedent and consequent pattern familiar to many human melodies, and we rarely hear choruses that are melodically memorable.  

{\bfseries Diversity:} Likelihood training encourages covering of all modes, so we expect the model to produce diverse samples.

{\bfseries -- \href{https://soundcloud.com/openai_audio/sets/jukebox-samples-re-renditions/s-IsBDzuVrO44}{Re-renditions}:} We generate multiple samples conditioned on artist and lyrics combinations that exist in our training data. While occasionally drum and bass lines or melodic intervals echo the original versions, we find that none of the generated samples is noticeably similar to the original songs.

We also generate multiple songs conditioned on the same artist and lyrics as Sample 1 to obtain Samples 9--12. All five sound interesting in their own ways with different moods and melodies with Sample 10 playing a harmonic at 00:14 as part of a blues riff, showing that the model has learned a wide range of singing and playing styles.

{\bfseries -- \href{https://soundcloud.com/openai_audio/sets/jukebox-samples-novel/s-OCmVIfH4il8}{Completions}:} We prime the model with 12 seconds of existing songs and ask it to complete them in the same styles. When the priming samples include singing, the continuations are more likely to imitate the original tunes and rhythms. Songs primed with more generic or common intros tend to be more diverse. Even generated samples that are close to the originals early on deviate completely into new musical material after about 30 seconds.

Re-renditions and completions are interesting and diverse, but overall, there is still room for improvement in music quality compared to the original songs.

{\bfseries -- \href{https://soundcloud.com/openai_audio/sets/jukebox-samples-full-tree/s-wbPtTR5KNh5}{Full tree}:} To understand diversity in a more systematic way, we generate multiple continuations from the same segment. We start with a one-minute sample and independently sample four times per one-minute extension. By the three minute mark, there are 16 completions. We can think of this branching tree as exploring different possibilities obtained by ancestral sampling. In the generated songs in the link, we hear diversity in singing and development even when the same initial segment is used. We note that this particular sample follows the lyrics more successfully than many.  For certain genres like hip hop and rap, where linearly moving the window does not yield good lyrics alignment, the chance of obtaining plausible singing is lower.

{\bfseries Novelty:} With the ability to condition on various styles, lyrics, and raw audio, we would like Jukebox to be a useful tool for both professional musicians and music enthusiasts alike. In this section, we are interested in exploring capabilities and applications of Jukebox.

{\bfseries -- \href{https://soundcloud.com/openai_audio/sets/jukebox-samples-novel-styles/s-SMgMBHByEVd}{Novel styles}:} We generate songs in an unusual genre typically not associated with an artist. In general, we find that it is fairly difficult to generalize to a novel style of singing while using the same voice as the artist embedding overpowers other information. In Joe Bonamassa and Frank Sinatra samples, we hear a modest variation in instrumentation, energy, and ambience depending on the genre embedding. However, our attempts to mix country singer Alan Jackson with unusual genres like hip hop and punk did not seem to move the samples away from a country style in meaningful ways.

{\bfseries -- \href{https://soundcloud.com/openai_audio/sets/jukebox-samples-novel-voice/s-Erfshq53w9W}{Novel voices}:} 
We pick artists whose voices are reproduced reasonably well by the model, and interpolate their style embeddings to synthesize new voices. Some blending, for instance, between Frank Sinatra and Alan Jackson in \href{https://soundcloud.com/openai_audio/0-5classic-pop-0-37889042/s-BAgcrk0m4dB?in=openai_audio/sets/jukebox-samples-novel-voice/s-Erfshq53w9W}{Sample 4}, still sounds similar to Frank Sinatra. In most cases, the model renders in a vaguely recognizable but distinct voice that preserves different vocal attributes. Samples \href{https://soundcloud.com/openai_audio/0-5country-0-5pop-in-140945998/s-JxRvdu6wQ6H?in=openai_audio/sets/jukebox-samples-novel-voice/s-Erfshq53w9W}{1} and \href{https://soundcloud.com/openai_audio/0-5pop-0-5jazz-in-the-style-9/s-EdlwU8CJuuK?in=openai_audio/sets/jukebox-samples-novel-voice/s-Erfshq53w9W}{2} conditioned on the C\'eline Dion embeddings divided by two have slightly different timbre and tone but capture her unique vibrato.

We also experiment with changing the style embedding in the middle of a song to create a duet (\href{
https://soundcloud.com/openai_audio/jukebox-duet-853523231/s-RcCBkn4iln7
}
{Sample 7}). This is another way of guiding generation during sampling. Continuing in another voice works best when the segment ends in an interlude; otherwise, the model blends voices in the middle of a word or a sentence.

{\bfseries -- \href{https://soundcloud.com/openai_audio/sets/jukebox-samples-novel-lyrics/s-qc1XhCOSjLw}{Novel lyrics}:} We ask Jukebox to sing poems and novel verses generated by GPT-2 \cite{gpt2} to demonstrate that it can indeed sing new lyrics. While the training data consists of song lyrics with limited vocabulary and constrained structure, the model has learned to follow along most prompts and sing even new words that are reasonably pronounceable (including technical terms from the deep learning literature). To get the best results, however, we find that it is useful to spell out difficult words or acronyms as they are spoken.  The generations are noticeably higher quality if the text matches the distribution of lyrics for the given artist, both in terms of length, and of rhyming or rhythmic qualities. For example, hip hop lyrics tend to be longer than most other genres, and the commonly emphasized syllables easily form clear rhythms.

{\bfseries -- \href{https://soundcloud.com/openai_audio/sets/jukebox-samples-novel-riffs/s-lo81x4FZFs2}{Novel riffs}:} Another useful application of Jukebox is the ability to record an incomplete idea and explore various continuations without ever needing to tabulate in symbolic representations, which would lose details of timbre and mood. We curate recordings of novel riffs by our in-house musicians and prime the model during sampling. Sample 6 starts with a musical style not widely used in Elton John's songs. The model still carries out the tune and develops it further. Similarly, the beginning of Sample 1 is a progressive jazz piece with a 5/4 polymeter, which has never been used in hip hop. Despite this novelty, the rhythm persists throughout the song and is incorporated naturally with rapping.

\subsection{VQ-VAE Ablations}
\begin{figure*}[h!]
\centering
\includegraphics[width=\textwidth]{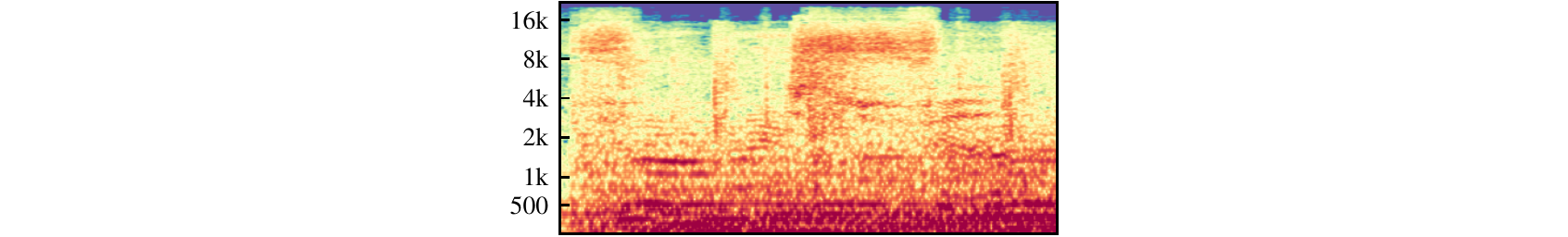} \\
\includegraphics[width=\textwidth]{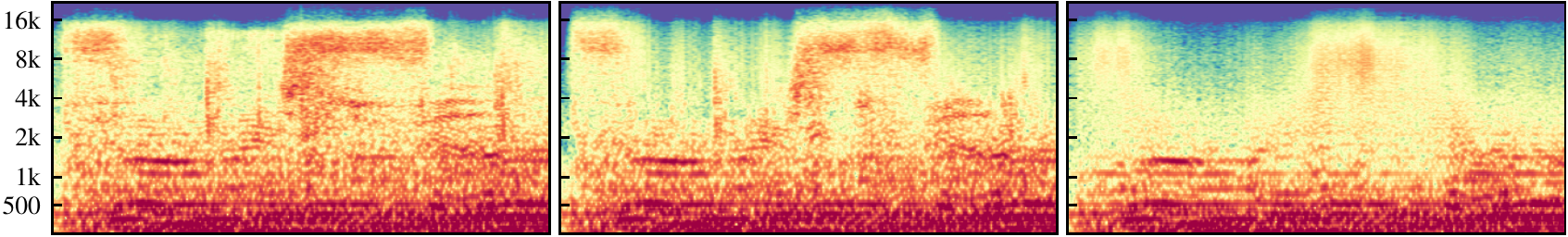} \\
\includegraphics[width=\textwidth]{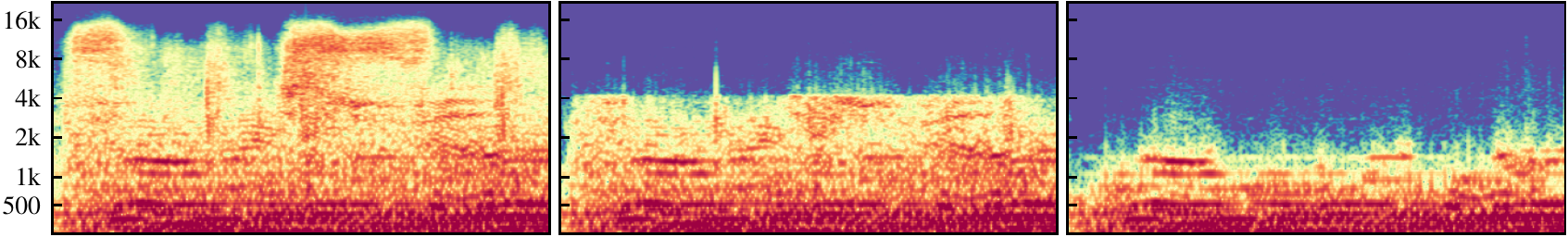} \\
\includegraphics[width=\textwidth]{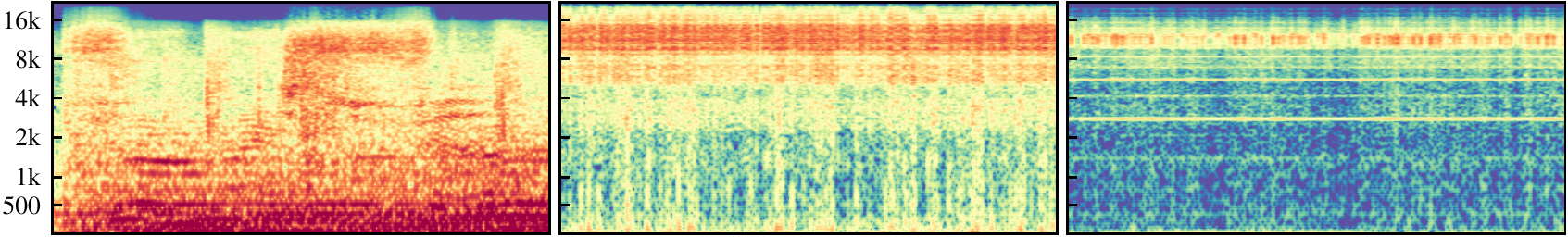} \\
\includegraphics[width=\textwidth]{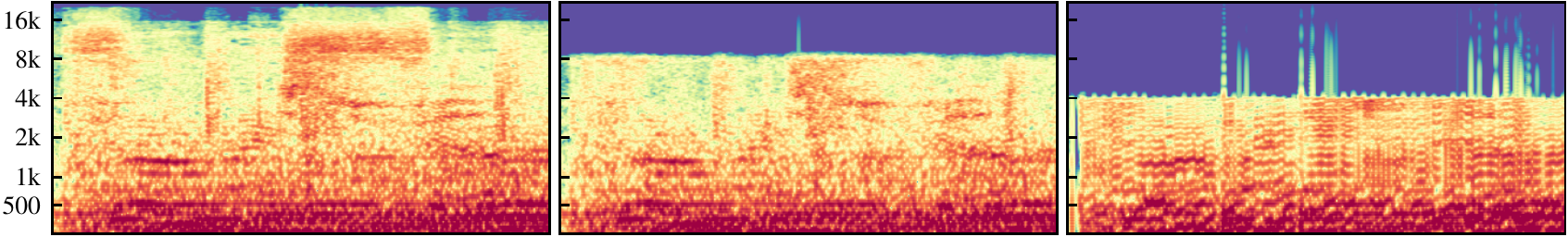} 
\caption{Comparison of reconstructions from different VQ-VAEs, x-axis is time and y-axis is frequency. The columns from left to right are bottom-, middle-, and top-level reconstructions at hop lengths 8, 32, and 128 respectively, visualized as Mel spectrograms. The first row is the ground-truth, and the second row shows the spectrograms of audio outputs from our VQ-VAE. In the third row, we remove the spectral loss, and see that the middle and top level lose high-frequency information. In the fourth row, we use a hierarchical VQ-VAE \cite{vqvae2} instead of separate auto-encoders (Figure \ref{fig:architecture:vqvae}), and we see the middle and top levels are not used for encoding pertinent information. Finally, the fifth row shows a baseline with the Opus codec that encodes audio at constant bitrates comparable to our VQ-VAE. It also fails to capture higher frequencies and adds noticeable artifacts at the highest level of compression.}
\label{fig:vq:spectrogram}
\end{figure*}

\begin{table}[ht]
\centering
\begin{tabular}{cccc}
\toprule
 & & \multicolumn{2}{c}{Spectral convergence (dB)} \\ 
Level & Hop length & Without restart & With restart \\
\midrule
Bottom &   8 & $-21.1$ & $-23.0$\\ 
Middle &  32 & $-12.4$ &  $-12.4$\\ 
Top    & 128 & $ -8.3$ & $-8.3$ \\ 
\bottomrule
\end{tabular}
\caption{Reconstruction fidelity degrades with higher compression. Restarting dead codes near random encoder outputs mitigates learning suboptimal codes.}
\label{tab:vq:compression}
\end{table}

\begin{figure}[h!]
\centering
\includegraphics[width=\columnwidth]{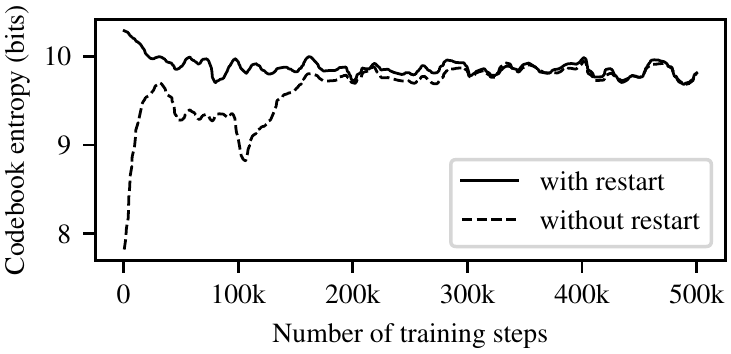}
\caption{Entropy of codebook with 2048 codes, i.e 11 bits, over training. Reviving dead codes near random encoder outputs ensures good codebook utilization from the start of training.}
\label{fig:vq:restart}
\end{figure}

\begin{table}[ht]
\centering
\begin{tabular}{cc}
\toprule
Codebook size & Spectral convergence (dB) \\
\midrule
 256 & $-15.9$ \\
2048 & $-23.0$ \\
No quantization & $-40.5$ \\
\bottomrule
\end{tabular}
\caption{Bottom-level VQ-VAE reconstruction results with different codebook sizes. Using larger codebooks helps reconstruction because it allows more information to be encoded at the bottleneck layers. Removing the bottleneck entirely yields almost perfect reconstruction.}
\label{tab:vq:codebook}
\end{table}

\begin{table}[ht]
\centering
\begin{tabular}{lc}
\toprule
Ablation & Spectral convergence (dB) \\
\midrule
None & $-8.3$ \\
Without spectral loss           & $-6.3$ \\
With single autoencoder    & $ \phantom{+}2.9$ \\
\bottomrule
\end{tabular}
\caption{Top-level codes are generally difficult to train well without spectral loss or with a single hierarchical autoencoder. Resulting reconstructions may lose some to most of information.}
\label{tab:vq:top-level}
\end{table}

We compare raw audio VQ-VAEs when trained with varying compression ratios, objectives, and architectures. 
As we use nonautoregressive decoders with continuous representation for output, we report spectral convergence \cite{specconv}, which measures the amount of spectral error relative to signal, as test error and proxy for reconstruction fidelity. 
We evaluate on 5000 held-out 3-second audio segments and report the average in decibels. 
All models in this section are trained with a batch size of 32, 3-second audio clips sampled at 44 kHz. 
As before, we use hop lengths of 8, 32, and 128 for the bottom, middle and top level respectively. 

In Table \ref{tab:vq:compression}, we see that increasing the hop size results in higher reconstruction error. Figure \ref{fig:vq:spectrogram} indeed shows that a significant amount of information, especially higher frequencies, is missing at middle and top levels across all ablations we ran. This is expected as audio is compressed more with larger hop sizes. 
To mitigate codebook collapse, we restart dead codes near random encoder embeddings. In Figure \ref{fig:vq:restart}, we see that this yields higher codebook usage even from early on in training. Models trained without random restarts can converge to the same test error and codebook usage but require more training steps. With poor initialization, these models sometimes end up with suboptimal codes hurting reconstruction fidelity.

Codebook size also matters, as it sets a limit on channel capacity through the bottleneck layers.
In Table \ref{tab:vq:codebook}, we find that reconstruction error increases considerably when the codebook size is reduced from 2048 to 256. We also compare with a model that uses continuous representations without vector quantization. We can think of this model as using a vastly large codebook with all encoder embeddings. This achieves almost perfect reconstruction with negligible spectral error.

When the model is trained with L2 loss only, reconstructions tend to sound muddy from missing high frequencies, and this problem is exacerbated as hop size is increased. In Figure \ref{fig:vq:spectrogram}, we see that top-level codes trained without spectral loss do not capture much information beyond 2 kHz, and obtain worse reconstructions (Table \ref{tab:vq:top-level}). However, we observe that while spectral loss helps encode more information, it also adds distortion artifacts which we hear as scratchy noise.

Lastly, we train a raw audio hierarchical VQ-VAE \cite{vqvae2} and find that it is generally difficult to push information to higher levels. This model is trained twice as long as the previous models, but middle and top-level reconstructions as shown in Figure \ref{fig:vq:spectrogram} are not capturing much. It is possible that higher level codes may have collapsed before bottom level starts to reconstruct the audio well. Making the bottom layers explicitly model residuals pushed more information to the top. But, we found separate autoencoders to be cleaner and more effective.

\section{Related Work} 
{\bfseries Generative modeling in deep learning:}
Generative models aim to learn the distribution of data by either explicitly by modeling the distribution or implicitly by constructing means to sample from it \cite{gantutorial}.
Modeling the interdependency within high-dimensional data was traditionally considered extremely difficult, but starting with Deep Boltzmann Machines \cite{dbm}, various kinds of deep generative models have been introduced.
Generative Adversarial Networks (GANs) \cite{gan} use generator and discriminator networks that contest each other to make the generated samples as indistinguishable as possible from the data, and they are renowned for their ability to generate high-quality pictures \cite{stylegan,biggan}.
Autoregressive generative models such as NADE \cite{nade}, PixelCNN \cite{pixelcnn}, and Transformers \cite{aiayn} use the chain rule of probability to factorize the joint distribution of data into a product of simpler distributions, and flow-based models \cite{nice,realnvp,flow,glow} learn a series of invertible transformations that maps the data distribution with a simpler one such as a Gaussian distribution.
Autoregressive flows \cite{maf,iaf} combine the two ideas to achieve faster density estimation or data generation.
Variational autoencoders (VAEs) \cite{rezende-vae,vae} impose a Gaussian prior on the latent code in an encoder-decoder setup from which data can be sampled.

{\bfseries Generative models for music:}
Generative modeling of symbolic music dates back to more than half a century, when \citet{hiller} introduced the first computer-generated music based on Markov chains.
There exists a variety of earlier approaches using rule-based systems \cite{rulebased}, chaos and self-similarity \cite{chaos}, cellular automata \cite{automata}, concatenative synthesis \cite{concat-music-synth}, and constraint programming \cite{constraint}.
More recent data-driven approaches include DeepBach \cite{deepbach} and Coconet \cite{coconet} which use Gibbs sampling to produce notes in the style of Bach chorals, MidiNet \cite{midinet} and MuseGAN \cite{musegan} which use generative adversarial networks, MusicVAE \cite{musicvae} and HRNN \cite{hrnn} which use hierarchical recurrent networks, and Music Transformer \cite{musictransformer} and MuseNet \cite{musenet} which use Transformers to autoregressively predict MIDI note events. 
There also have been a number of approaches for synthesizing music conditioned on symbolic music information, such as NSynth \cite{nsynth} which uses WaveNet-style autoencoder, Mel2Mel \cite{mel2mel} and Wave2Midi2Wave \cite{maestro} which synthesize music using WaveNet conditioned on a piano roll representation, and GanSynth \cite{gansynth} which uses generative adversarial networks to produce magnitude spectrograms together with instananeous frequencies for easier spectrogram inversion.
Generative models for music can also be used for music style transfer, as seen in Midi-VAE \cite{midivae} which uses a variational autoencoder to transfer styles between classical and jazz music, LakhNES \cite{lakhnes} which uses a Transformer architecture to generate chiptune music, and Universal Music Translator Network \cite{umtn} which uses a denoising autoencoder that can disentangle musical style and content.

{\bfseries Sample-level generation of audio:}
In recent years, a variety of generative models for raw audio have been introduced.
WaveNet \cite{wavenet} performs autoregressive sample-by-sample probabilistic modeling of raw waveform using a series of dilated convolutions to exponentially increase the context length. It can produce realistic audio either unconditionally or by conditioning on acoustic features or spectrograms. The autoregressive nature of WaveNet makes the sampling notoriously slow, and it uses a categorical distribution for audio samples which introduces quantization noise.
Parallel WaveNet \cite{parallelwavenet} improves upon this by instead using a mixture of logistics distribution, a continuous probability distribution, and performing probability density distillation which learns a parallel feed-forward network from a pre-trained autoregressive model, allowing faster sampling of high fidelity audio. ClariNet \cite{clarinet} achieves similar audio quality using a simple Gaussian distribution instead and thus having a closed-form loss function, eliminating the need for Monte-Carlo sampling.
SampleRNN \cite{samplernn} uses a multi-scale, hierarchical recurrent neural network with convolutional upsampling to model long-range complex structures.
WaveRNN \cite{wavernn} uses recurrent neural networks that operate separately on the most significant and the least significant bytes, which can be efficiently deployed in mobile devices while having comparable audio quality to WaveNet.
WaveGlow \cite{waveglow} is a flow-based model for parallel sample-level audio synthesis, which can be trained with a straightforward maximum-likelihood estimation and thus is advantageous to the two-stage training process needed for distillation.
Parallel WaveGAN \cite{parallelwavegan} and MelGAN \cite{melgan} are GAN-based approaches directly modeling audio waveforms, achieving similar quality as WaveNet and WaveGlow models with significantly fewer parameters.
While the approaches above serve as sophisticated generative models for raw audio to be conditioned on a compact and controllable representation of audio such as Mel spectrograms, MelNet \cite{melnet} takes a different approach of hierarchically generating accurate high-resolution Mel spectrograms, after which a simple gradient-based optimization can produce high-fidelity audio.

{\bfseries VQ-VAE:}
\citet{vqvae} introduced VQ-VAE, an approach of downsampling extremely long context inputs to a shorter-length discrete latent encoding using a vector quantization, and they showed that it can generate both high-quality images and audio, as well as learn unsupervized representations of phonemes. \citet{vqvae2} extended the above model by introducing a hierarchy of discrete representations for images and showed that the resulting model can learn to separate high-level semantics into the highest level of discrete codes which have the largest receptive field, while capturing local features like textures in the lower levels with smaller receptive fields. They used the hierarchical model to generate high-diversity and high-fidelity images for the conditional ImageNet and FFHQ datasets. \citet{sanderschallenge} tried variants of this approach where instead of a single encoder there are successive encoders that each further compress the lossy discrete encodings from the previous levels. A downside of this approach is that information is lost at each step and requires separate training for each VQ-VAE level, and it leads to a hierarchy collapse problem. \citet{sandershierarchical} used AR decoders which are known to cause the problem of ignoring the latent variables, and they suggested ways to mitigate it. The feed-forward decoders from \cite{vqvae2} do not suffer from this issue, and thus we use their approach.

{\bfseries Speech synthesis:}
Producing natural human voice entails an understanding of linguistic features, mapping of sounds, and steerability of expression. Many text-to-speech (TTS) systems rely on highly engineered features \cite{formant-tts}, carefully curated sound segments \cite{unit-selection}, statistical parametric modeling \cite{hmm-tts}, and often complex pipelines as described in \cite{dv1}. These approaches are fairly involved and produce unnatural or inarticulate voices. More recent works like Deep Voice 3 \cite{dv3}, Tacotron 2 \cite{tt2}, and Char2Wav \cite{c2w} learn speech synthesis end-to-end using sequence-to-sequence architecture \cite{seq2seq}. The design space is vast,
but in general, typical approaches comprise of a bidirectional encoder, a decoder, and a vocoder to build text representations, audio features, and the final raw waveforms. To generate multiple voices, text-to-speech models can also condition on the speaker identity \cite{wavenet,dv2,multispeaker-tts} as well as text prompt. By learning and manipulating auxiliary embeddings, models can mimic a new voice \cite{voice-cloning,voice-loop} at test time. These methods, however, require labeled data. Ideas like clustering \cite{i-vector}, priming \cite{style-tokens}, and variational autoencoders \cite{vae-tts-1,vae-tts-2} have been used to learn broader styles of speech and control expressivity in an unsupervised way. There are also works on synthesizing singing by additionally controlling pitch and timbre. Similar to TTS literature, early works use concatenative methods \cite{concat-singing-synth} that join short segments of curated singing, and statistical parametric methods \cite{hmm-singing-1,hmm-singing-2} which allow modeling of timbre from training data. Both approaches impose fairly strong assumptions resulting in noticeable artifacts. \cite{neural-singing-synth} train a neural TTS model with a parametric vocoder to separate pitch and timbre which can be controlled at generation time.

\section{Future work}
While our approach represents a step forward in the ability to generate coherent long raw audio music samples, we recognize several directions for future work. Great music generation should be high quality over all time scales: it should have a developing musical and emotional structure across the entire piece, local notes and harmonies that always make sense, nuanced and appropriate small timbral and textural details, and audio recording quality that balances and blends the multiple voices well, and without unwanted noise.  We view our current model as stronger on the mid-range time scales: often the model generates samples that locally sound very good, with interesting and diverse harmonies, rhythms, instruments, and voices. We have frequently been very impressed how the melody and rhythm generated suits a particular lyric extremely well.  However, while the samples stay consistent over longer time scales, we notice they don't have traditional larger music structures (such as choruses that repeat, or melodies that have a question and answer form).   Additionally, on the smallest scale, we sometimes hear audio noise or scratchiness.  

Beyond the quality of the samples, we also would look to diversify the languages and styles the model is able to generate.  Our current model has been trained only on songs whose primary language as detected by \cite{cld2} is English.  In the future, we would look to include other languages and artists.  We believe this will be of interest both for generating strictly in those styles, and because historically we have seen much creativity and development coming from unusual blends of existing musical styles.

Finally, we consider it very important that computer music generation also serves as a tool for human musicians, and increasingly those interested in music but without formal training. While we are able to steer our current model somewhat through lyric and midi conditioning, we can imagine many other possible ways for humans to influence the generations, including indicating the mood or dynamic at various sections, or controlling when drums, singers, or other instruments should play.  

The current model takes around an hour to generate 1 minute of top level tokens. 
The upsampling process is very slow, as it proceeds sequentially through the sample. Currently it takes around 8 hours to upsample one minute of top level tokens.  We can create a human-in-the-loop co-composition process at the top level only, using the VQ-VAE decoders to get a fast upsampling of the top level tokens to hear a very rough sense of what the model generates. The top-level model generates multiple samples, the person picks a favorite (listening to the rough VQ-VAE decoding), and then the model continues generating multiple samples continuing the favorite. This process would be significantly improved with faster generation and Transformer upsampling steps. Our models have fast parallel evaluation of likelihood but slow autoregressive sampling. We can instead use a model with fast parallel sampling but slow autoregressive likelihood evaluation \cite{iaf}, and distill the information from our current model into it \cite{parallelwavenet}. The distillation works by generating samples from the parallel sampler and evaluating it likelihood and entropy using the parallel likelihood evaluator, and then optimising the sampler by minimising the KL divergence of it from our current model. 

\section{Conclusion}
We have introduced Jukebox, a model that generates raw audio music imitating many different styles and artists. We can condition this music on specific artists and genres, and can optionally specify the lyrics for the sample. We laid out the details necessary to train a Hierarchical VQ-VAE to compress the music effectively into tokens. While previous work has generated raw audio music in the 20--30 second range, our model is capable of generating pieces that are multiple minutes long, and with recognizable singing in natural-sounding voices.

\section{Acknowledgement}
We would like to thank John Schulman and Will Guss for producing and performing novel riffs for our sampling experiments, and Rewon Child, Aditya Ramesh, Ryan Lowe and Jack Clark for providing feedback for initial drafts of this paper. 
\bibliography{main}

\bibliographystyle{icml2020}

\clearpage
\pagebreak

\appendix

\section{Scalable Transformer}
\label{scalabletransformer}
    We make the Sparse Transformer \cite{sparsetransformer} more scalable and easier to implement by a few small changes. We implement a simpler attention pattern that has the same performance without needing custom kernels to implement. We simplify the initialization by using the same initalization scale in the whole model without rescaling the weights based on fan-in and depth, and we optimize the memory footprint with fully half-precision training, i.e. storing the model weights, gradients and the optimizer states in half precision and performing computations in half precision as well. To cope with the narrower dynamic range of the fp16 format, we use dynamic scaling of the gradient and Adam optimizer states. 

{\bfseries Axis-aligned attention patterns:} The Sparse Transformer \cite{sparsetransformer} sparsifies the attention pattern by reshaping the input sequence into a 2-D sequence of shape $($blocks, block length$)$ to use factorized attention. They observe that the strided attention pattern works best for images and audio because it does not have the state bottleneck of the fixed attention. However, their implementation require specialized CUDA kernels. We can obtain a similar pattern by doing masked row, masked column, and unmasked previous-row attention. While the masked row captures the local context, the masked column and unmasked previous-row attention captures the context of all previous rows. We observe the same computational speed as well as training loss with this pattern. Each of these can be implemented directly as a dense attention by transposing or slicing the input sequence along appropriate axes, and thus do not require special CUDA kernels to implement. This can be easily extended to video too. Complementary to our work, a similar pattern was introduced in \cite{axial} where they also used axis-aligned attention but instead used a two-stream architecture. \\

\begin{figure}[t]
    \centering
    \begin{subfigure}{\columnwidth}
        \includegraphics[width=\columnwidth]{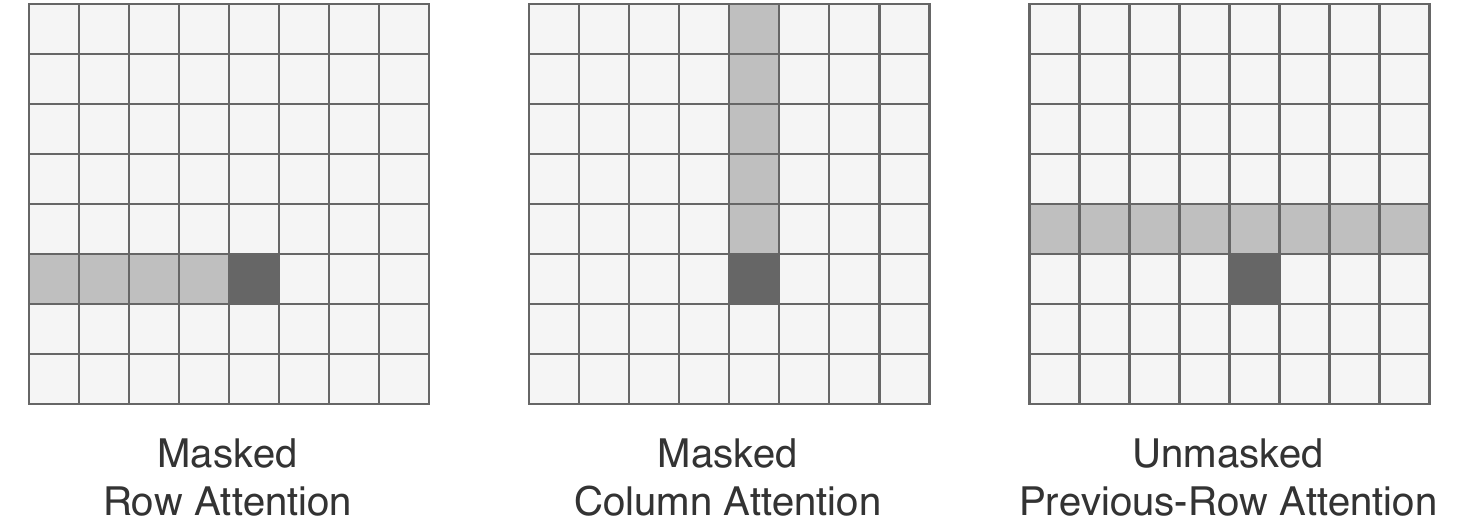}
        \caption{Three axis-aligned attention patterns are sparse attention patterns that allow autoregressive generative modeling while only using simple Python-level array manipulation. Masked row and column attention patterns use autoregressive masks, whereas unmasked previous-row attention is fully visible.}
        \label{fig:architecture:attention-patterns}
    \end{subfigure}\\[1.5em]
    \begin{subfigure}{\columnwidth}
        \includegraphics[width=\columnwidth]{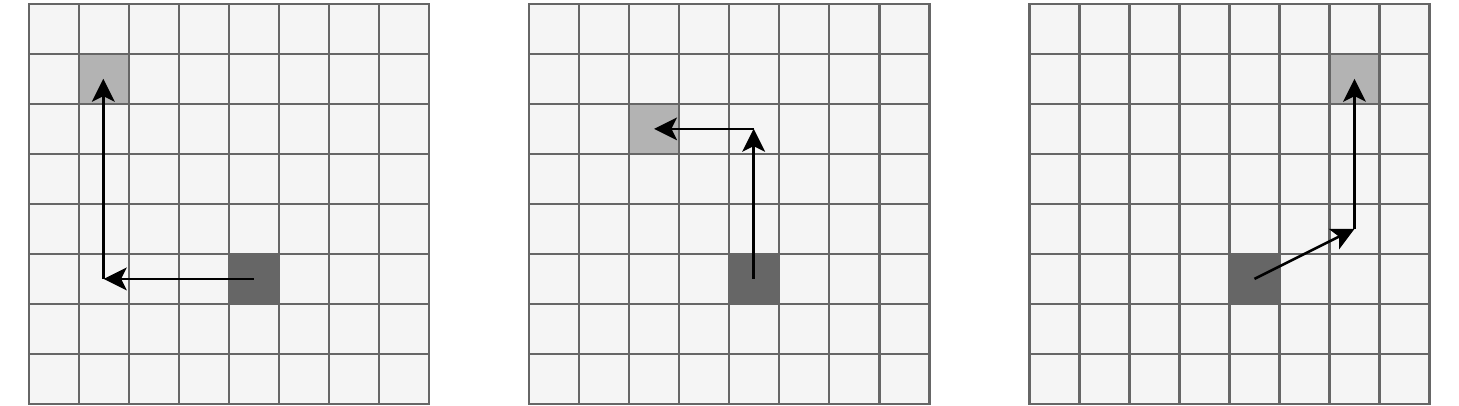}
        \caption{Combining two of the attention patterns, each position can attend to any of the previous positions, while not causing a state bottleneck as in fixed sparse attention \cite{sparsetransformer}.}
        \label{fig:architecture:attention-hops}
    \end{subfigure}
    \caption{Axis-aligned attention patterns}
    \label{fig:architecture:factorized-attention}
    \vspace{1em}
\end{figure}

{\bfseries Half-precision parameters and optimizer state with dynamic scaling:} To allow training large models, \cite{sparsetransformer} uses recompute with gradient checkpointing, performs computations using half precision activations and gradients, and uses dynamic loss scaling. While this speeds up training on Volta cores, one still has a high memory usage from storing the parameters and Adam state in full float precision. To scale our models further, we store our matmul parameters and their Adam state in half precision, thus halving our memory usage. We use a single parameter $s$ to set the scale of all weights and initialize all matmul and input/output embeddings\footnote{We share the input and output embedding} to $N(0, s)$, and position embeddings to $N(0, 2s)$. The initialization ensures all parameters are in a similar dynamic range, and allows us to train in half precision completely without loss in training performance. For the Adam state tensors \texttt{(\lstinline{m_t}, \lstinline{v_t})} we do dynamic scaling. For each iteration and for every parameter, we rescale its state tensors before casting so that their maximum corresponds to the maximum value of the float16 range, thus maximizing the use of the float16 range. Thus, we store the state \texttt{\lstinline{m_t}} as the tuple \texttt{(scale, (\lstinline{m_t}/scale).half())}, where \texttt{scale = \lstinline{m_t}.max()/float16.max()}, and similarly for \texttt{\lstinline{v_t}}. The above lets us fit models of size $1$B parameters into memory for our large context of 8192 tokens. To train even larger models, we use GPipe \cite{gpipe}.

\vfill
\pagebreak
\section{Experimental details}

\subsection{Music VQ-VAE}\label{sec:music-vqvae}

\begin{figure}[t]
    \centering
    \begin{subfigure}{\columnwidth}
        \centering
        \includegraphics[width=\columnwidth]{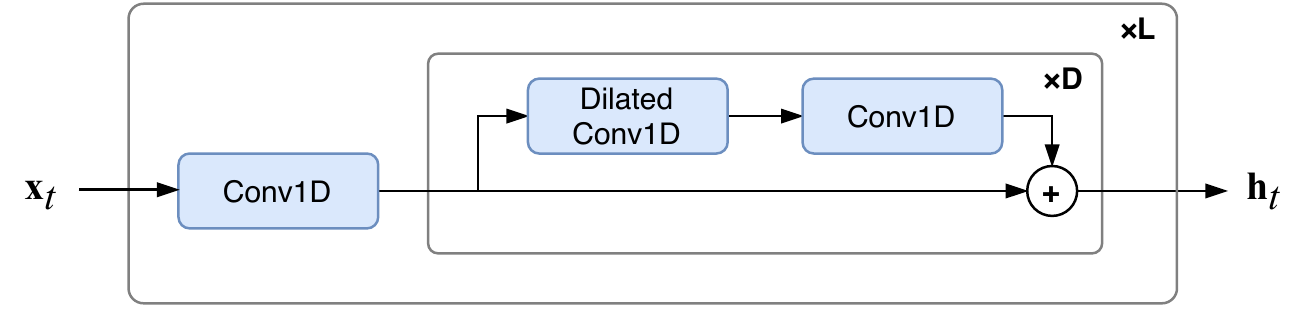}
        \caption{
        The encoder compresses the raw audio input into a sequence of embeddings. The length of this latent representation relative to the raw audio duration determines the amount of compression, and is an important factor for the trade-off between fidelity and coherence.
        }
        \label{fig:architecture:vqvae-encoder}
    \end{subfigure}
    \\[1em]
    \begin{subfigure}{\columnwidth}
        \centering
        \includegraphics[width=\columnwidth]{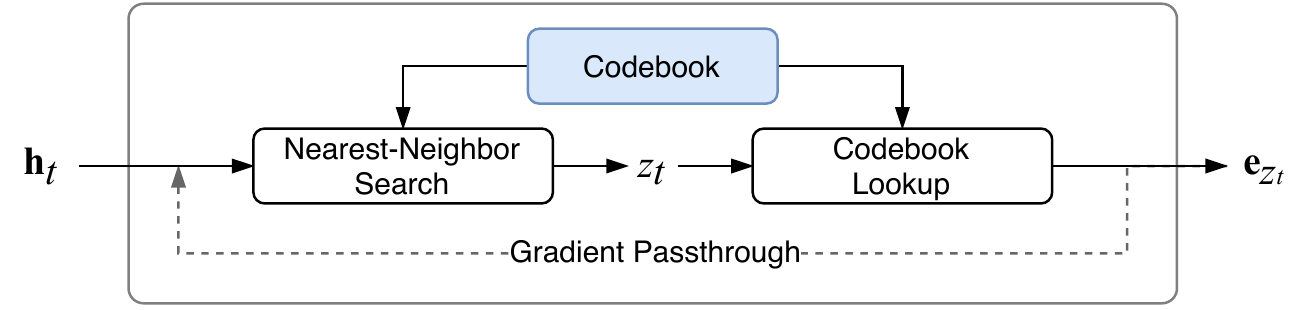}
        \caption{The bottleneck takes the sequence of embeddings from the encoder and maps it into a sequence of code vectors from the codebook. This sequence of code indices is used as a discrete representation to be modeled by the priors. Larger codebooks improve fidelity but may be more difficult to compress.}
        \label{fig:architecture:vqvae-bottleneck}
    \end{subfigure}
    \\[1em]
    \begin{subfigure}{\columnwidth}
        \centering
        \includegraphics[width=\columnwidth]{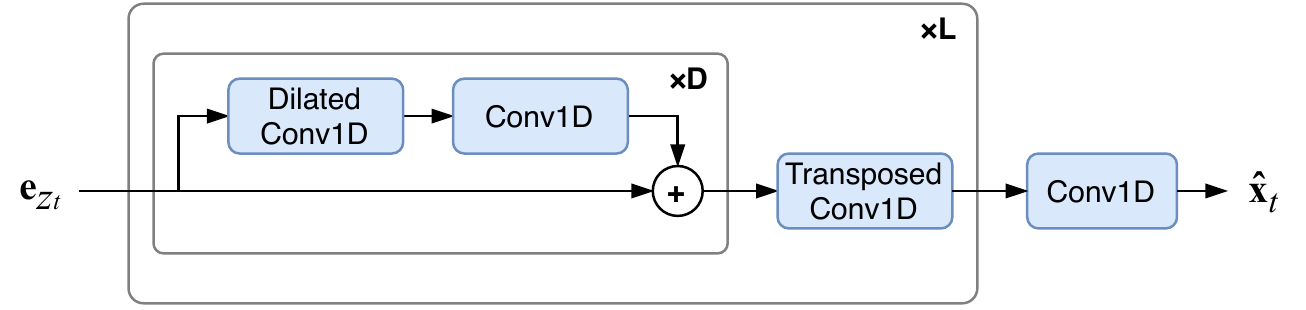}
        \caption{The decoder reconstructs the raw audio from latent representations. It is a mirror of the encoder where dilations constracts by a factor of 3 down to 1 at the last block.
        The final Conv1D projects to the desired number of audio channels and also acts as a smoothing operation after a sequence of transposed convolutions.}
        \label{fig:architecture:vqvae-decoder}
    \end{subfigure}
    \caption{Components of the VQ-VAE model}
    \label{fig:architecture:vqvae-components}
    \vspace{1em}
\end{figure}

We have three separate raw audio VQ-VAEs to produce discrete codes at varying hop sizes for the bottom, middle, and top priors.
All autoencoders comprise non-causal, dilated 1-D convolutions, and are trained independently using non-autoregressive reconstruction losses. Basic building blocks in these networks share the same architecture, as shown in Figure \ref{fig:architecture:vqvae-components}. Each encoder block consists of a downsampling convolution, a residual network, and a 1D convolution with a kernel size of 3. Dilation is grown by a factor of 3 in these residual networks to increase the receptive field. The decoder block mirrors this exactly with a 1D convolution with the kernel size of 3, a residual network with dilation contracting across depth, and an upsampling transposed convolution. Here, all resampling convolutions use a kernel size of 4 and stride 2 so that each building block changes the hop length by a factor of 2. To get higher compression in time, we simply stack more of these blocks. For example, using seven blocks yields a hop length of 128 for the top level autoencoder.

Each residual network has four residual blocks in the middle and top VQ-VAEs resulting in a receptive field of 120 ms and 480 ms for the respective discrete tokens. Because increasing the residual depth helped improve reconstruction quality slightly, we doubled the number of residual blocks for the bottom level. This dramatically increases the receptive field to about 2 seconds per code but the actual receptive field is mostly local.

We also experimented with having a single decoder and modeling the residuals to separate out learned representations as in \cite{vqvae2}, hoping upsampling priors would simply fill in local musical structure. However, pushing information to the top level was quite challenging as the bottommost level reconstructs almost perfectly early on in training. When we add auxiliary objectives to encourage the top to be used more, the top-level codes add serious distortions to the final output. A similar challenge is shown in \cite{sanderschallenge}.

\subsection{Music Priors and Upsamplers}
\begin{figure*}[t]
    \begin{subfigure}{0.58\textwidth}
    \centering
    \includegraphics[height=16em]{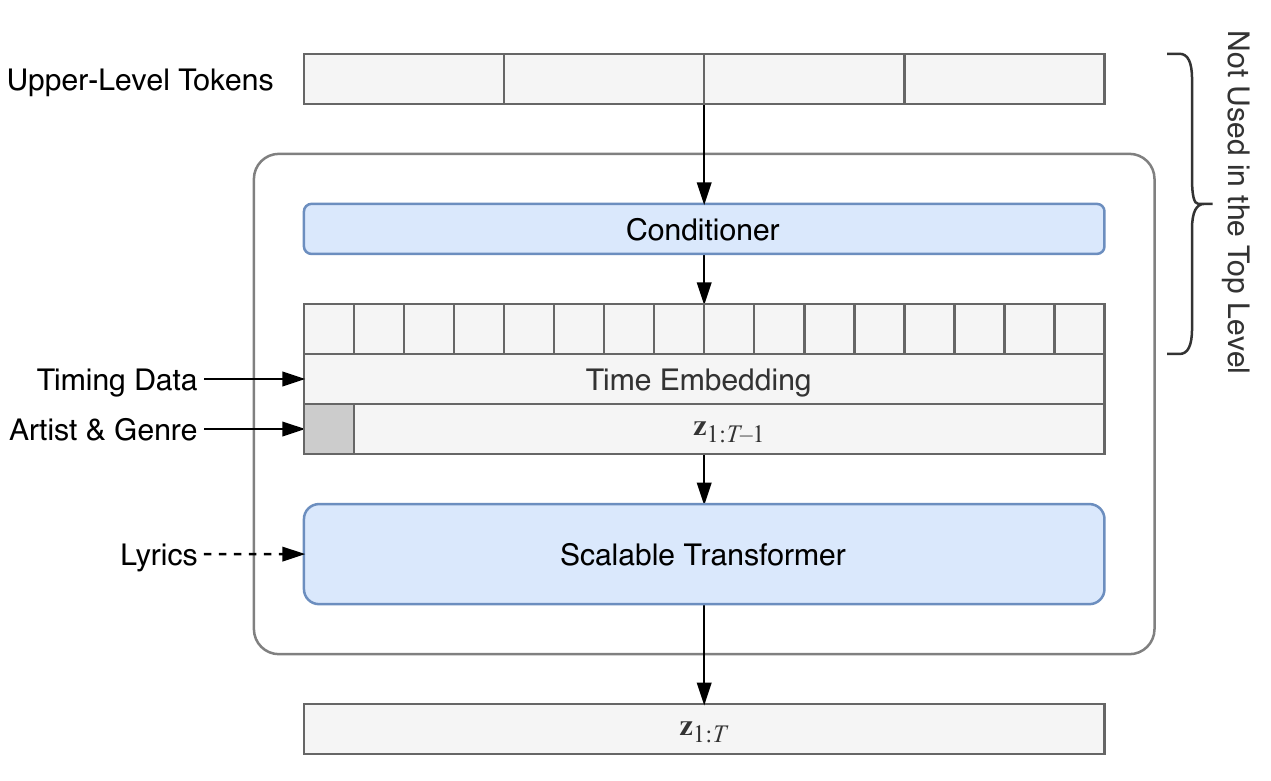}
    \caption{The structure of our prior models, performing next-token prediction at each level. The Transformer takes the embeddings of the tokens $\z_{1:T-1}$ prepended by the sum of the artist and genre embeddings, in addition to the time embedding that encodes relative and absolute timing of the segments in the duration of the song. The upsampler priors additionally take the tokens from the upper level, which are fed to the conditioner network and added to the input sequence. The top-level prior takes lyrics as conditioning information as well (see Figure \ref{fig:architecture:encdec}).
    }
    \label{fig:architecture:conditioning}
    \end{subfigure}\quad\quad
    \begin{subfigure}{0.38\textwidth}
    \centering
    \includegraphics[height=15em]{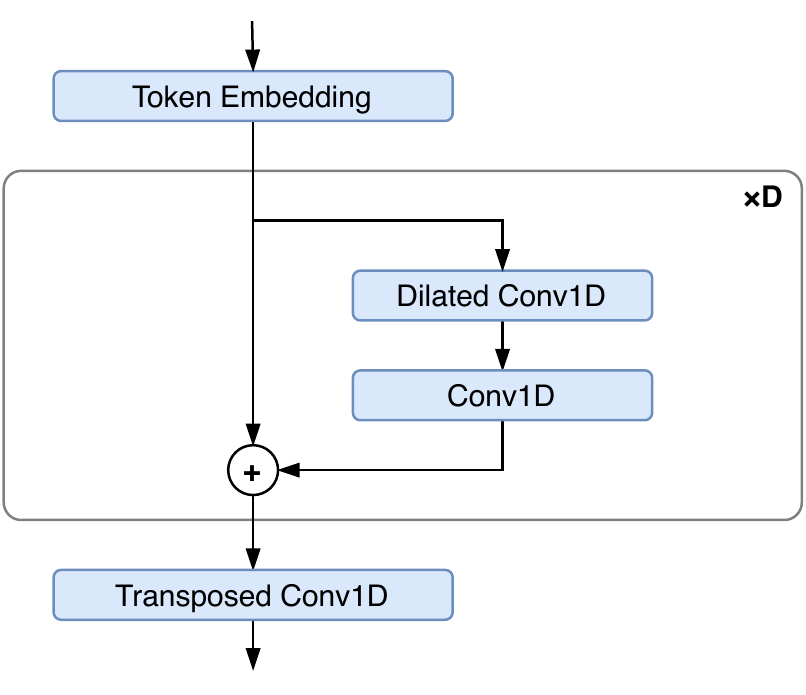}
    \caption{The conditioner network takes the tokens from the upper level, and their embedding vectors go through non-causal WaveNet-like layers with increasingly dilated convolutions. The transposed 1-D convolution upsamples the sequence to the higher temporal resolution of the current level.}
    \label{fig:architecture:conditioner}
    \end{subfigure}\\[2em]
    \begin{subfigure}{0.68\textwidth}
    \centering
    \includegraphics[height=29em]{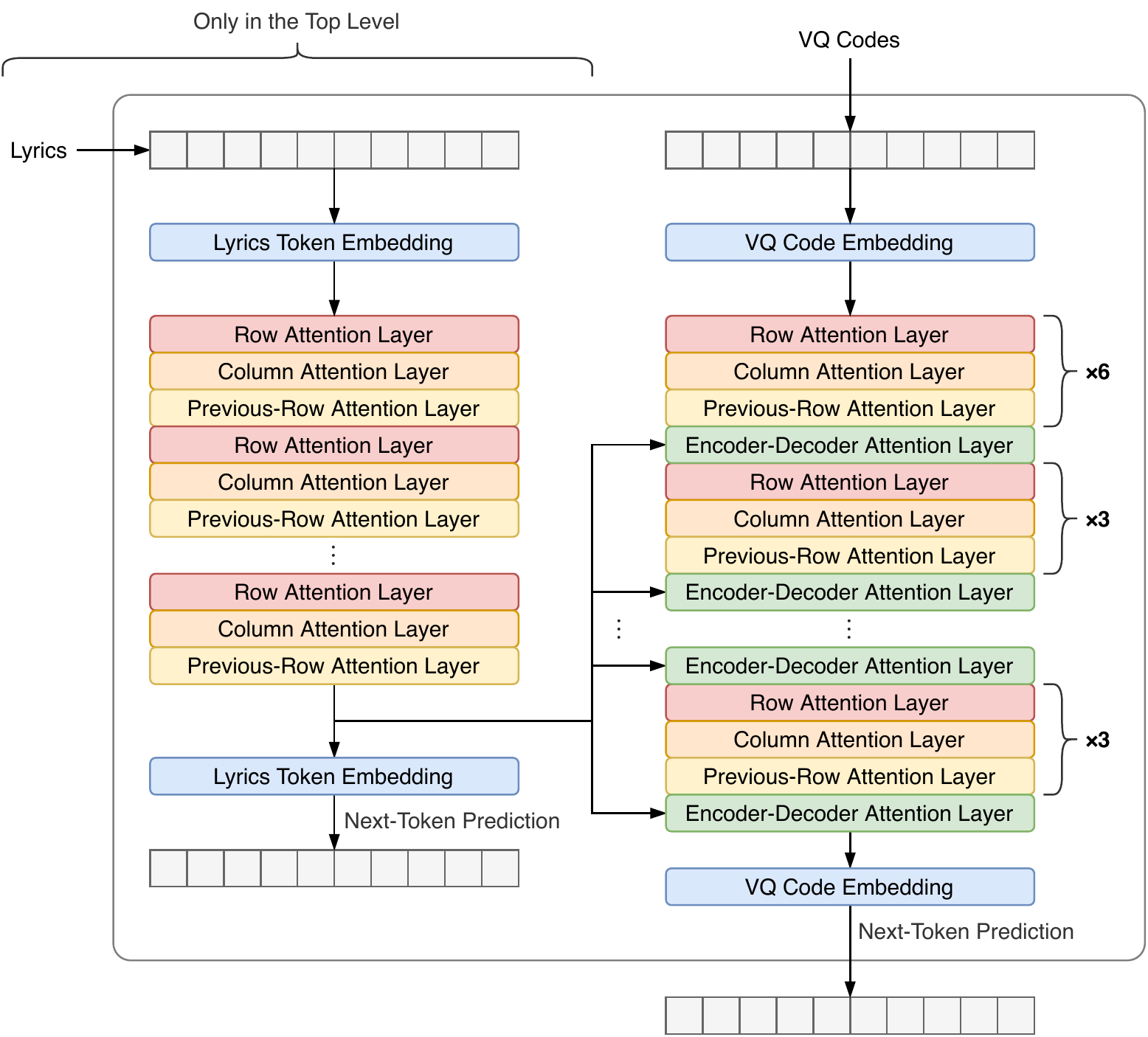}
    \caption{The Scalable Transformer architecture, shown with the lyrics Transformer used in the top-level prior. The Transformer layers use the three factorized attention types alternatingly, i.e. repeating row, column, and previous-row attentions. In the top-level prior, the VQ Transformer additionally includes interleaved encoder-decoder attention layers that apply lyrics conditioning by attending on the activation of the last encoder layer.}
    \label{fig:architecture:encdec}
    \end{subfigure}\quad\quad
    \begin{subfigure}{0.28\textwidth}
    \centering
    \vspace{2em}
    \includegraphics[height=22em]{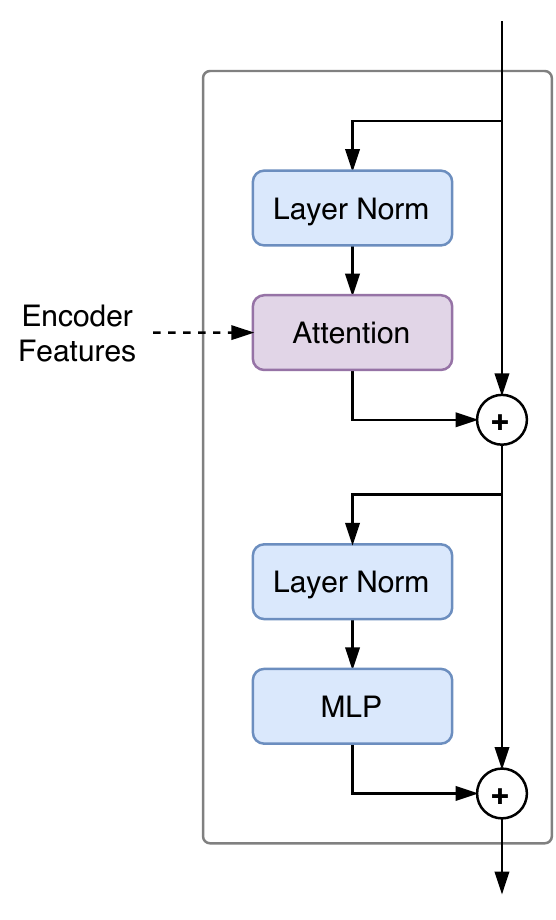}
    \vspace{1em}
    \caption{Each Transformer layer is a residual attention block, which performs two residual operations, attention and MLP, each prepended with layer normalization. Depending on the layer's type, it uses either one of the three factorized attentions or encoder-decoder attention taking the lyrics features from the encoder.}
    \label{fig:architecture:resblock}
    \end{subfigure}\quad\quad
    \caption{Detailed architecture of the music prior and upsampler models}
    \label{fig:architecture:prior}
\end{figure*}
Architectural details of our music prior and upsampler models are depicted in Figure \ref{fig:architecture:prior}. They perform autoregressive modeling of tokens at each level, conditioned on information such as artist and genre, as well as the tokens from the upper level in the case of the upsamplers (Figure \ref{fig:architecture:conditioning}).
Each artist and genre are learned as embedding vectors, whose sum is provided as the very first token in each sequence.
In addition, positional embedding is learned as a function of each position's absolute and relative timing in the duration of the song.
In upsampler models, upper-level tokens are upsampled by the conditioner network, using WaveNet-style dilated convolutions followed by a transposed 1-D convolutional layer (Figure \ref{fig:architecture:conditioner}).

When the model is trained on lyrics, the top-level prior takes lyrics data corresponding to each audio segment and uses them to train an encoder-decoder Transformer as shown in Figure \ref{fig:architecture:encdec}. 
All transformer stacks use sparse self-attention layers with the three factorized attention types (row, column, and previous-row) repeating, and encoder-decoder attention layers, when present, are interleaved with the other attention types.
Each layer consists of residual connections of an attention and an MLP feedforward network, each prepended by layer normalization  (see Figure \ref{fig:architecture:resblock}).

\FloatBarrier

\subsection{Hyperparameters}
For all Transformers' residual blocks, we use MLP blocks with the same width as the model width, and attention blocks with queries, keys, and values with width 0.25 times the model width. For all convolutional residual blocks, we use convolutions with same channels as the model width. 

\label{sec:hps}
\begin{table}[ht!]
\centering
\begin{tabular}{l|c} 
    \toprule
    Sample rate & 44100 \\
    Sample length & 393216 \\
    Hop lengths & 8, 32, 128  \\
    Embedding width & 64 \\
    Residual block width & 64, 32, 32 \\
    Residual blocks (per 2x downsample) & 8, 4, 4\\
    Conv filter size & 3 \\
    Conv channels & 32 \\
    Dilation growth rate & 3 \\
    Commit weight $\beta$ & 0.02 \\
    Codebook EMA $\gamma$ & 0.99 \\
    Codebook size & 2048 \\
    Spectral loss STFT bins & 2048, 1024, 512  \\
    Spectral loss STFT hop length & 240, 120, 50 \\
    Spectral loss STFT window size & 1200, 600, 240 \\
    Initialization scale & 0.02 \\
    Batch size & 256 \\
    Training steps & 384618 \\
    Learning rate & 0.0003 \\
    \bottomrule
\end{tabular}
\caption{VQ-VAE hyperparameters}
\label{tab:vqvae-hps}
\end{table}

\begin{table}[ht!]
\centering
\begin{tabular}{l|c}
    \toprule
    & 1B upsamplers \\
    \midrule
    Sample length & 262144, 65536 \\
    Context length & 8192 \\
    Transformer width & 1920 \\
    Transformer layers & 72 \\
    Attention heads & 1 \\ 
    Factorized attention shape & (128, 64) \\
    Conditioner residual block width & 1024 \\
    Conditioner residual blocks & 16 \\
    Conditioner conv filter size & 3 \\ 
    Conditioner conv channels & 1024 \\
    Conditioner dilation growth rate & 3 \\
    Conditioner dilation cycle & 8 \\
    Initialization scale & 0.004, 0.008\\
    Batch size & 192, 184 \\
    Training steps & 265000, 279000 \\
    Learning rate & 0.0003 \\
    Adam $\beta_2$ & 0.95 \\
    Weight decay & 0.01 \\ 
    \bottomrule
\end{tabular}
\caption{Middle- and bottom-level upsampler hyperparameters}
\label{tab:upsamp-hps}
\end{table}

\begin{table}[ht!]
\centering
\begin{tabular}{l|c}
    \toprule
    & 5B prior \\
    \midrule
    Sample length & 1048576 \\
    Context length & 8192 \\
    Transformer width & 4800 \\
    Transformer self-attention layers & 72 \\
    Attention heads & 8 \\ 
    Factorized attention shape & (128, 64)  \\
    Lyrics encoder tokens & 512 \\
    Lyrics encoder width & 1280 \\ 
    Lyrics encoder layers & 18 \\
    Lyrics encoder attention heads & 4 \\
    Lyrics encoder factored attention shape & (32, 16) \\
    Encoder-Decoder attention layers & 7 \\
    Initialization scale & 0.002 \\
    Encoder initialization scale & 0.014 \\
    Batch size & 512 \\
    Training steps & 310500 \\
    Learning rate & 0.00015 \\
    Adam $\beta_2$ & 0.925 \\
    Weight decay & 0.002 \\
    \bottomrule
\end{tabular}
\caption{Top-level prior hyperparameters}
\label{tab:top-hps}
\end{table}

\clearpage
\subsection{$t$-SNE Plot of Artists}

\vspace{2em}

\begin{figure}[h!]
    \begin{minipage}{\textwidth}
    \centering
    \includegraphics[width=\textwidth]{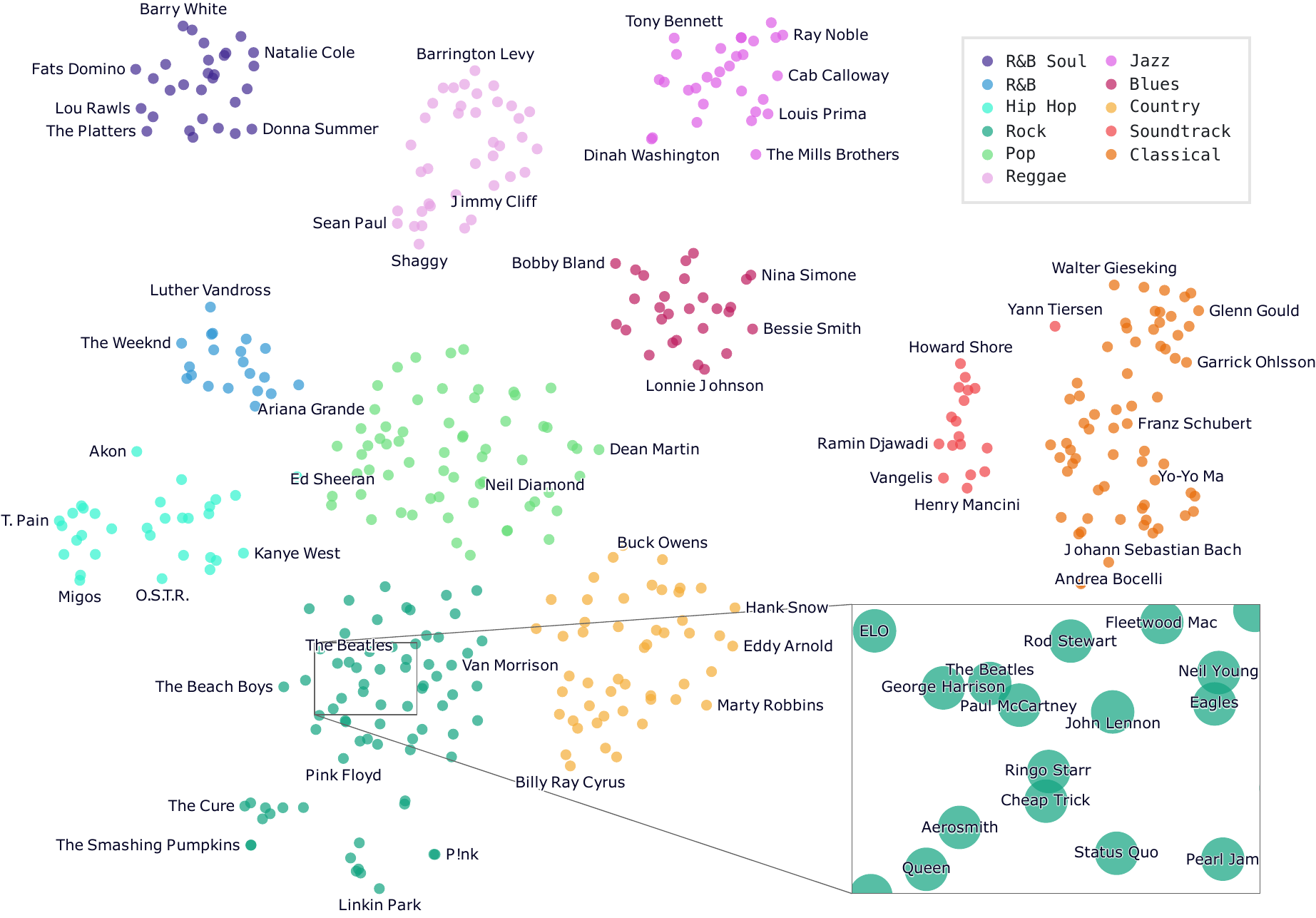}
    \caption{$t$-SNE of (artist, genre) embedding. The overall clustering shows very clearly how genres are related to one another. The broadest of all, pop, is situated in the middle of rock, country, blues, hip hop, and many more. Soundtrack and classical form their own island. Within a genre, we see a similar trend among artists. John Lennon, Paul McCartney, George Harrison and Ringo Starr are clustered around The Beatles. Cheap Trick which has a number of Beatles covers is also found near. Because we are showing only about 400 artists here, not all neighboring artists may be related. For an interactive version, we point to our \href{https://openai.com/blog/jukebox}{blog post}.
    }
    \label{fig:tsne}
    \end{minipage}
\end{figure}

\end{document}

%% file: header/preamble.tex
\usepackage{microtype}
\usepackage{graphicx}
\usepackage{booktabs} 

\usepackage[utf8]{inputenc}
\usepackage{amsfonts}
\usepackage{amsmath}
\usepackage{amsthm}
\usepackage{amssymb}
\usepackage{subcaption}
\usepackage{placeins}
\usepackage{todonotes}
\usepackage{caption}
\usepackage{subcaption}
\usepackage{dblfloatfix}
\usepackage[bottom]{footmisc}

\usepackage[linkcolor=blue]{hyperref}

\usepackage[T1]{fontenc}
\usepackage{listings}
\usepackage{pgfplots}

%% file: header/math.tex
\newcommand{\Loss}{\mathcal{L}}
\newcommand{\x}{\mathbf{x}}
\newcommand{\h}{\mathbf{h}}
\newcommand{\z}{\mathbf{z}}
\newcommand{\e}{\mathbf{e}}
\newcommand{\sg}{\mathrm{sg}}

\newcommand{\abs}[1]{|#1|}

\newcommand{\Norm}[1]{\left\lVert#1\right\rVert}
\DeclareMathOperator*{\smallsum}{\textstyle\sum}

%% file: header/defns.tex
\newcommand{\recons}{{\text{recons}}}
\newcommand{\commit}{{\text{commit}}}
\newcommand{\codebook}{{\text{codebook}}}
\newcommand{\spec}{{\text{spec}}}
\newcommand{\stft}{{\text{STFT}}}

\newcommand\hiddenfootnote[1]{%
  \begingroup
  \renewcommand\thefootnote{}\footnote{#1}%
  \endgroup
}